\documentclass[useAMS,usenatbib]{mnras}

\usepackage[english]{babel}
\usepackage{amsmath}
\usepackage{amssymb}
\usepackage{txfonts}
\usepackage{graphicx}
\usepackage{color}

\newcommand{\pd}{\partial}
\newcommand{\ud}{\ensuremath{\mathrm{d}}}

\title{Multi-D Simulations of Ultra-Stripped Supernovae to Shock Breakout}

\author[B.~M\"uller et al.]
  {Bernhard~M\"uller,$^{1,2}$\thanks{E-mail: bernhard.mueller@monash.edu}
  Daniel~W.~Gay,$^{2,1}$
  Alexander~Heger$,^{1,3}$
  Thomas~M.~Tauris,$^{4,5,6}$
  and 
\newauthor
Stuart~A.~Sim$^{2}$
\\
$^{1}${Monash Centre for Astrophysics, School of
  Physics and Astronomy, Monash University, Victoria
  3800, Australia} \\
$^{2}${Astrophysics Research Centre, School
  of Mathematics and Physics, Queen's University
  Belfast, Belfast, BT7~1NN, United Kingdom} \\
$^{3}${Tsung-Dao Lee Institute, Shanghai 200240, China.} \\
$^{4}$Argelander-Institut f\"ur Astronomie, Universit\"at Bonn, Auf dem H\"ugel 71, 53121 Bonn, Germany\\
$^{5}$Max-Planck-Institut f\"ur Radioastronomie, Auf dem H\"ugel 69, 53121 Bonn, Germany\\
$^{6}$Department of Physics and Astronomy, Aarhus University, Ny Munkegade 120, 8000 Aarhus C, Denmark
}

\begin{document}

\label{firstpage}
\pagerange{\pageref{firstpage}--\pageref{lastpage}}

\maketitle

\begin{abstract}
The recent discoveries of many double neutron star systems and their
detection as LIGO-Virgo merger events call for a detailed
understanding of their origin.  Explosions of ultra-stripped stars in
binary systems have been shown to play a key role in this context and
have also generated interest as a potential explanation for rapidly
evolving hydrogen-free transients. Here we present the first attempt
to model such explosions based on binary evolution calculations that
follow the mass transfer to the companion to obtain a
consistent core-envelope structure as needed for reliable predictions
of the supernova transient.  We simulate the explosion in 2D and
3D, and confirm the modest explosion energies
$\mathord{\sim} 10^{50}\,\mathrm{erg}$ and small kick velocities
reported earlier in 2D models based on bare carbon-oxygen
cores.  The spin-up of the neutron star by asymmetric accretion is
small in 3D with no indication of spin-kick
alignment. Simulations up to shock breakout show the mixing of
sizeable amounts of iron group material into the helium envelope. In
view of recent ideas for a mixing-length treatment (MLT) of
Rayleigh-Taylor instabilities in supernovae, we perform a detailed
analysis of the mixing, which reveals evidence for buoyancy-drag
balance, but otherwise does not support the MLT
approximation. The mixing may have implications for the spectroscopic
signatures of ultra-stripped supernovae that need to be investigated
in the future. Our stellar evolution calculation also predicts
presupernova mass loss due to an off-centre silicon deflagration
flash, which suggests that supernovae from extremely stripped cores
may show signs of interactions with circumstellar material.
\end{abstract}


\section{Introduction}
\label{sec:intro}
In recent years, there has been considerable progress in our
theoretical understanding of both the explosions of massive stars as
core-collapse supernovae and the modelling of their progenitors in close binary systems. 
In light of the increasing number of discoveries of double neutron star (DNS) systems
and their mergers, leading to the high-frequency gravitational wave (GW) bursts detected by LIGO/Virgo \citep[GW170817,][]{abbott_17},
it is of utmost importance to understand the formation of DNS systems \citep{tauris_17}.
In particular, the second supernova (SN) explosion is a key ingredient to gain further knowledge of 
the survival rates of such systems and thereby the expected LIGO-Virgo detection rates.

The second SN also determines the kinematics of the surviving DNS systems 
(i.e., their resulting systemic runaway velocity) which is important for determining the offset distance from their host galaxies 
by the time the two NSs merge and produce a short gamma-ray burst \citep{fb13,aaa+17d,bbf+17} and a kilonova \citep{kn14,sha+17,cfk+17,aaa+17e,scj+17,dps+17}
following the GW signal.

The vast majority (if not all) of massive stars are born in binary systems \citep{sdd+12} and it
is well known that the outcome of stellar evolution in close binaries differs significantly from that of single stars \citep{lan12}. 
Binary interactions affect, for example, the rotation rate, the amount of envelope material, and the final core mass prior to the core collapse \citep{bhl+01,plp+04}.
In extreme scenarios for the second SN forming a close-orbit DNS system,
it has been demonstrated \citep{tauris_13,tauris_15} that the progenitor stars 
are ultra-stripped prior to their core collapse; i.e., these stars basically become almost naked metal cores 
due to mass transfer via so-called Case~BB Roche-lobe overflow (RLO) to the first-born NS of the system.  
Therefore, to make further progress in our understanding of the formation of close-orbit DNS systems 
and the LIGO-Virgo GW sources, it is  timely and necessary to start modelling 
such ultra-stripped SN explosions in detail. 

On the side of transient observations, the advent of high-cadence
surveys (e.g., PanSTARRS, \citealp{chambers_16}; PTF, \citealp{law_09}; iPTF, \citealp{mlr+17}; 
SkyMapper, \citealp{keller_07}) has increased the interest 
in studying the diversity  among core-collapse SN events
\citep[e.g.,][]{drout_11,pejcha_15c,terreran_17}.
Especially faster and fainter transients become more accessible by
observations. In terms of
the explosion parameters, this means that the regime of 
ultra-stripped SNe with small ejecta
masses and explosion energies comes
into focus thanks to the
new observational capabilities. There
are already noteworthy cases like SN~2005ek
and SN~2010X \citep{drout_13,tauris_13,moriya_17} that
have been identified as candidates for ultra-stripped SNe,
and more are likely to follow. It is imperative to put
the interpretation of such fast and faint transients on
a firmer footing using self-consistent explosion models.

In this paper, we therefore investigate the low-mass 
end of stripped-envelope progenitors of Type~Ib/c SNe, i.e.,
ultra-stripped SNe, by means of multi-dimensional
simulations, extending earlier work by \citet{suwa_15}.
Different from the 2D study of \citet{suwa_15}, we conduct
simulations both in 3D and 2D, thus further
extending the growing list of successful 3D explosion models
in the field
\citep{takiwaki_12,takiwaki_14,melson_15a,melson_15b,lentz_15,mueller_15b,mueller_16b,janka_16,mueller_17,summa_17,roberts_17,chan_17,ott_17,kuroda_18}.
Moreover, we perform calculations from collapse to shock breakout
and  apply, for the first time in a 3D SN simulation, 
a progenitor  obtained from binary stellar
evolution modelling with  a helium envelope rather than 
 exploding bare C/O cores.

Although the 2D study of \citet{suwa_15}
established the basic parameters of ultra-stripped explosions
from a theoretical point of view, namely a low explosion energy
of $\mathord{\sim}10^{50}\,\mathrm{erg}$, a small nickel mass of
$\mathord{\lesssim}\,0.01 M_\odot$, and a small kick velocity, a
number of questions about this SN channel still remain open,
and our approach allows us to address some of these.
On a very basic note, 3D modelling is necessary simply for confirming
the 2D results of
\citet{suwa_15} and for modelling
the possible spin-up of neutron star (NS) due to asymmetric accretion. 

Perhaps more importantly, the connection
between the theoretical models of \citet{suwa_15} and
observed fast and faint transients still remains rather tenuous.
That the theoretically predicted explosion properties of
ultra-stripped models
roughly fit the light curves of such a type~Ic event
like SN~2005ek 
has been demonstrated
by \citet{moriya_17} using 1D models, but there
is only rough agreement between the observationally inferred
nickel masses
of $0.01-0.05\;M_{\odot}$
and explosion energies of a few $10^{50}\, \mathrm{erg}$
\citep{drout_11,moriya_17} and the theoretical predictions, which
may partly be due to degeneracies in the light curve fits.
An unambiguous identification of ultra-stripped
SNe needs to be based on spectroscopy. 

This, however, requires an understanding of the mixing by Rayleigh-Taylor
instabilities during the propagation of the shock through the envelope.
The extent of the mixing is crucial as even the gross spectral
type (Ib vs.\ Ic) is sensitive to the mixing of radioactive nickel
into the helium envelope. This is because gamma-ray energy deposition from 
the nickel affects the non-thermal excitation of helium, as shown by
\citet{dessart_12,dessart_15}
and \citet{hachinger_12} in detailed non-LTE 
radiative transfer simulations and by \citet{piro_14} based on 
analytic estimates.  The question of mixing in Type~Ib/c SNe 
is in fact relevant not only for
ultra-stripped progenitors, but for the 
entire class of hydrogen-free progenitors, and has so far been
explored only to a very limited extent even in parameterised 2D models
\citep{hachisu_91,hachisu_94,kifonidis_03} --- in contrast to the very
extensive body of computational studies on mixing in Type~II
SNe such as SN~1987A
\citep{arnett_89,benz_90,mueller_91,fryxell_91,hachisu_92,kifonidis_00,kifonidis_03,hammer_10,ellinger_13,wongwathanarat_15}
and Cas~A \citep{wongwathanarat_17}. By combining self-consistent
multi-dimensional explosion models and a realistic envelope
structure from binary evolution, we can now start to address
the question of mixing in Type~Ib/c SNe more reliably and take
a first step towards connecting the multi-D explosion
simulations to observations by detailed spectral modelling in
the future.

Our paper is organised as follows:  In Section~\ref{sec:setup},
we describe the binary progenitor model and the numerical methods
and setup used for multi-dimensional simulations of the
explosion from collapse to shock breakout.
We then discuss the results of the SN
simulations in Sections~\ref{sec:second}
and \ref{sec:envelope}. Section~\ref{sec:second} focuses
on the explosion properties -- i.e., the explosion energy,
the composition of the inner ejecta, and the PNS
mass, spin, and kick. Mixing instabilities in the envelope are
addressed in Section~\ref{sec:envelope}, where we
analyse the growth conditions for Rayleigh-Taylor-driven
mixing and describe the final state of mixing at shock breakout.
We also investigate to what extent the non-linear phase of
the Rayleigh-Taylor instability can be described by
effective 1D models in the vein of mixing-length theory, 
as recently suggested by \citet{duffell_16}
and \citet{paxton_17}.
Section~\ref{sec:applications} discusses
our results in the context of DNS system
properties.
We conclude with a summary and a discussion of the
implications of our results in
Section~\ref{sec:conclusions}.

\section{Input Models and Numerical Methods}
\label{sec:setup}

\subsection{Progenitor Model}\label{subsec:progenitor}
\citet{tauris_15} calculated a large grid of progenitor models for ultra-stripped
SNe by evolving helium stars of metallicity $Z=0.02$ including
mass transfer via Case~BB\footnote{Strictly speaking: Case~BA, Case~BB or Case~BC depending on 
whether the RLO is initiated while the helium star undergoes core helium burning, helium shell burning
or has evolved to core carbon burning or beyond.} RLO to a NS companion. 
This phase of binary evolution \citep[e.g.,][]{tv06} follows after the 
high-mass X-ray binary stage which evolves into common-envelope evolution
where the hydrogen envelope is ejected via in-spiral of the NS.
In the subsequent phase of evolution the exposed core (i.e., the helium star)
will initiate RLO to its NS companion if the orbit is not too wide.

\begin{figure}
\includegraphics[width=1.10\linewidth]{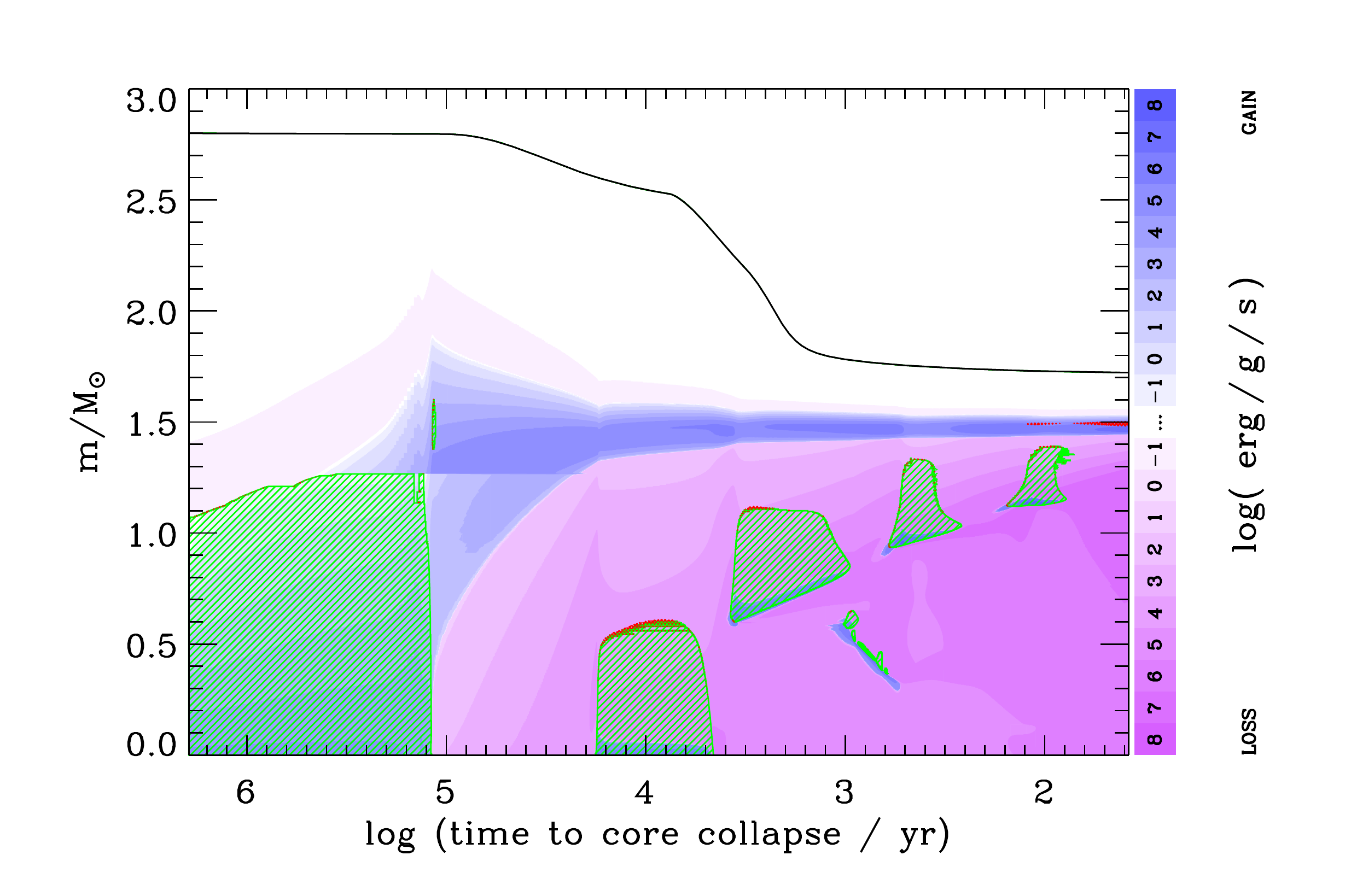}
\caption{Kippenhahn diagram of the $2.8\,M_{\odot}$ helium star undergoing Case~BC
RLO to its NS companion, as modelled by \citet{tauris_15}.
The plot shows cross-sections of the helium star in mass-coordinates
from the centre to the surface of the star, along the $y$-axis, as a function of stellar age on the $x$-axis.
The plotted age interval, for which we used the BEC code spanning a total time of $\mathord{\simeq} 1.95\,{\rm Myr}$, is  truncated here at $\mathord{\sim} 37.9\,{\rm yr}$ prior to core collapse (see text).  
The remaining evolution is modelled with the \textsc{Kepler} code.
Case BC~RLO is initiated during core carbon burning at $\log\, (t_*)\simeq 3.8$. 
The green hatched areas denote zones with convection; red colour indicates semi-convection.
The intensity of the blue/purple colour indicates the net energy-production rate.
\label{fig:Kipp}}
\end{figure}

In this paper, we consider the model with an initial helium star
mass of $2.8\,M_\odot$ and an initial orbital period of $20 \,
\mathrm{d}$. As a result of a stellar wind, the mass is reduced to $\mathord{\sim} 2.5\,M_{\odot}$ 
by the time the helium star initiates mass transfer while undergoing core carbon burning. 
The subsequent stage of Case~BC RLO reduces the helium star mass to $1.72\,M_{\odot}$ with a remaining helium envelope of $0.217\,M_\odot$ (Figure~\ref{fig:Kipp}).

Up to the stage of early oxygen burning, the evolution is followed using
the binary evolution code \textsc{BEC} of \citet{wellstein_01}, which
is based on the single-star code of \citet{langer_98}; for details see
Section~2 in \citet{tauris_15}. Progenitor rotation is not
explicitly considered in the calculation.
Due to tidal coupling one expects the progenitor to
spin extremely slowly with a spin
period of the order of the final
orbital period of $19.4 \, \mathrm{d}$.
In general, the final orbital periods
(and hence the pre-collapse spin
periods) for ultra-stripped SN progenitor
are expected to vary considerably, however,
with a broad distribution down to
less than $1\, \mathrm{h}$ \citep{tauris_15}.

At oxygen ignition, the binary has detached again, and the final C/O core mass of the helium star is about $\mathord{\sim}1.47\,M_\odot$.  Since the nuclear network in \textsc{BEC} is not well-suited for advanced phases well beyond carbon burning, we then map the model into the \textsc{Kepler} code during neon burning, which is ignited in a shell off-centre in this low-mass core.  This is similar to what has been done by \citet{heger_00} where a mapping was done when a central temperature of $10^9\,\mathrm{K}$ was reached.   \textsc{Kepler} has been well-developped to properly treat the advanced burning stages.  In particular, silicon burning is treated using a quasi-nuclear statistical equilibrium (QNSE) network, and the iron core past silicon burning uses a nuclear statistical equilibrium (NSE) network \citep{weaver_78,heger_10}.  The remaining time from mapping (off-centre neon ignition) to core collapse is $\sim37.9\,\mathrm{yr}$.

The subsequent evolution to collapse is noteworthy.  Other recent
works on ultra-stripped SNe
\citep{tauris_13,tauris_15,tauris_17,suwa_15} have already remarked
upon the structural similarities of the progenitor cores to
single-star electron-capture
supernova (ECSN) progenitors
\citet{nomoto_84,nomoto_87,jones_13,jones_14,doherty_17} and low-mass
iron-core progenitors \citet{woosley_15}, which are essentially due to
the small C/O core mass and result in similar explosion dynamics.  The
small C/O core mass also has other interesting consequences because
some of the final core and shell burning episodes occur under strongly
degenerate conditions \citep{woosley_15}, which can lead to
off-centre ignition and very violent flash-like burning that triggers
presupernova mass ejection.  In the progenitor considered here, off-centre neon burning first ignites at about $0.47\,M_\odot$ is accompanied by of-centre oxygen burning as the shell progresses further inward.  Various smaller of-centre O and Ne shell burning stages occur further our as well during that phase but have little effect in the over-all progress of the inward burning shell.  When the shell reaches about $0.09\,M_\odot$, silicon burning ignites violently causing a sound wave that travels to the surface, steepening into a shock as it runs donw the density gradient.  This leads to the ejection of most of the helium envelope. Only
a small residual envelope of $0.02\,M_\odot$ remains. In this work, we
cut the ejected matter and evolve the rest of the star further to
collapse, which occurs $78 \, \mathrm{d}$ later. At this stage, much
of the ejected material has already reached radii of
$\mathord{\sim}10^{15} \, \mathrm{cm}$, and expansion velocities are
of order $\mathord{\sim}1000 \, \mathrm{km} \, \mathrm{s}^{-1}$.

Due to the presupernova mass ejection, the observable transient may
be very strongly affected by interaction and evolve into a Type~Ibn
supernova at some stage.  It could thus appear considerably brighter
than the faint transients that have been predicted for ultra-stripped
progenitors \citep{moriya_17}. Based on the properties of the ejected
shell, we still expect to see a distinguishable supernova Type Ib/c-like transient
before the supernova stars to interact with the circumstellar material
from the pre-collapse mass ejection.  With maximum ejecta velocities
of $16,000 \, \mathrm{km} \, \mathrm{s}^{-1}$ in the supernova, we
expect strong interaction features to emerge no earlier than about
$12\, \mathrm{d}$ after the explosion, i.e., after the peak of the
light curve judging by the results of \citet{moriya_17} -- especially if
we consider that the mass of the supernova ejecta will be even lower
 than in \citet{moriya_17} so that the transient should evolve
more rapidly. For the first phase of the observable transient, one can
therefore justifiably disregard the circumstellar material when
discussing mixing, light curves and spectra, although the interaction
phase will be of great interest for future work.

\subsection{Simulating the Neutrino-Driven Explosion}
We simulate the collapse, the post-bounce accretion phase, and the
initial explosion phase using the neutrino hydrodynamics code
\textsc{CoCoNuT-FMT} \citep{mueller_15a}.  The hydrodynamics module
\textsc{CoCoNuT} \citep{dimmelmeier_02_a,mueller_10} solves the
equations of general relativistic hydrodynamics in spherical polar
coordinates in an unsplit finite-volume approach using piecewise
parabolic reconstruction \citep{colella_84} and the HLLC Riemann
solver \citep{mignone_05_a}. The metric equations are solved in the
extended conformal flatness approximation \citep{cordero_09}, and a
spherically symmetric metric is assumed.

As in previous works \citep{mueller_15b}, we employ a mesh-coarsening
scheme for variable resolution in the longitudinal direction to avoid
strong time step constraints near the grid axis, and we model the
inner region with density $\gtrsim 10^{11} \, \mathrm{g}\,
\mathrm{cm}^{-3}$ in spherical symmetry, using a mixing-length
treatment for proto-neutron star (PNS) convection.  In the high-density
regime, we use the equation of state of \citet{lattimer_91} with a
bulk incompressibility modulus of $K=220 \, \mathrm{MeV}$.

We conduct two 2D runs (s2.8-2D-a and s2.8-2D-b) and one 3D run
(s2.8-3D).  The only difference between the 2D models is that we have
run model s2.8-2D-b with 6th-order extremum preserving reonstruction
\citep{colella_08,sekora_09} instead of the standard piecewise
parabolic reconstruction.

\subsection{Simulation to Shock Breakout}

When the explosion energy
is reasonably converged and further energy input by neutrino heating
becomes negligible, we map models s2.8-3D and s2.8-2D-b into the Newtonian hydrodynamics
code \textsc{Prometheus} \citep{fryxell_91,mueller_91}.
\textsc{Prometheus} is a directionally-split implementation of the
piecewise-parabolic method of \citet{colella_84}. In 3D, we use an overset
grid Yin-Yang grid consisting of two spherical polar coordinate
patches \citep{kageyama_04,wongwathanarat_10a} as implemented by
\citet{melson_15a}.  The initial grid resolution on each patch is $1600 \times 56 \times 148$, corresponding to an angular resolution of $1.6^\circ$.
The initial radial grid is equally spaced in $\log r$ with a
a resolution of $ \delta r/r=6.9 \times 10^{-3}$.

\textsc{Prometheus} allows for a moving radial grid
\citep{mueller_94}; in principle an arbitrary
grid velocity function $\dot{r}$ can be specified. In our models,
we choose a grid velocity of
of the form
\begin{equation}
\label{eq:ugrid}
\dot{r}_i=(\alpha+\beta i) r_i,
\end{equation}
for the radial zone $i$.
This allows us to make the expansion of the grid non-homologous so
that the inner boundary can ``catch up'' with the outward-moving
ejecta once the central region becomes sufficiently evacuated. This
largely eliminates the need to remove interior grid zones to increase
the time step (as in \citealp{hammer_10,wongwathanarat_13}).  Due to
the form of~Equation~(\ref{eq:ugrid}), the grid retains equal spacing
in $\ln r$.\footnote{This is because
$\ud/\ud t \,(\ln r_{i+1}-\ln r_{i+1})=
\alpha+\beta (i+1)-\alpha -\beta i = \beta$ is independent of $i$.
}

In the \textsc{Prometheus} runs, we use the equation of state of
\citet{timmes_00}. Self-gravity is accounted for in the Newtonian
approximation; as for the \textsc{CoCoNuT} model, the monopole
approximation is employed.

\begin{figure}
\includegraphics[width=\linewidth]{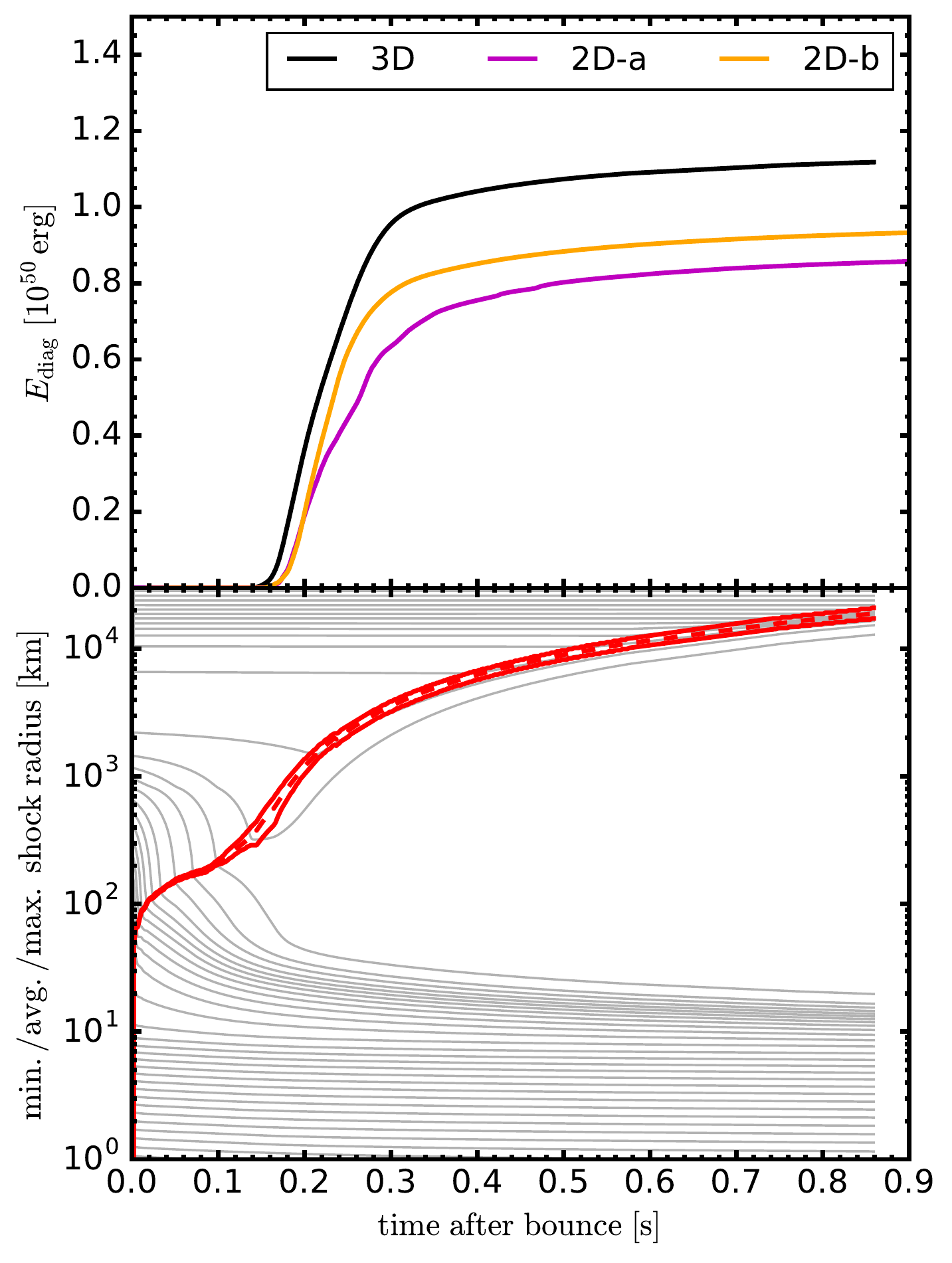}
\caption{Top: Evolution of the diagnostic explosion energy
for the 3D model (black) and the two axisymmetric models 
(magenta and orange). Bottom: Maximum, minimum (red solid
lines), and average (red, dashed) shock radius for
model s2.8-3D along with selected mass
shell trajectories (grey).
\label{fig:explosion}}
\end{figure}

\begin{figure*}
\centering
\includegraphics[width=0.49\linewidth]{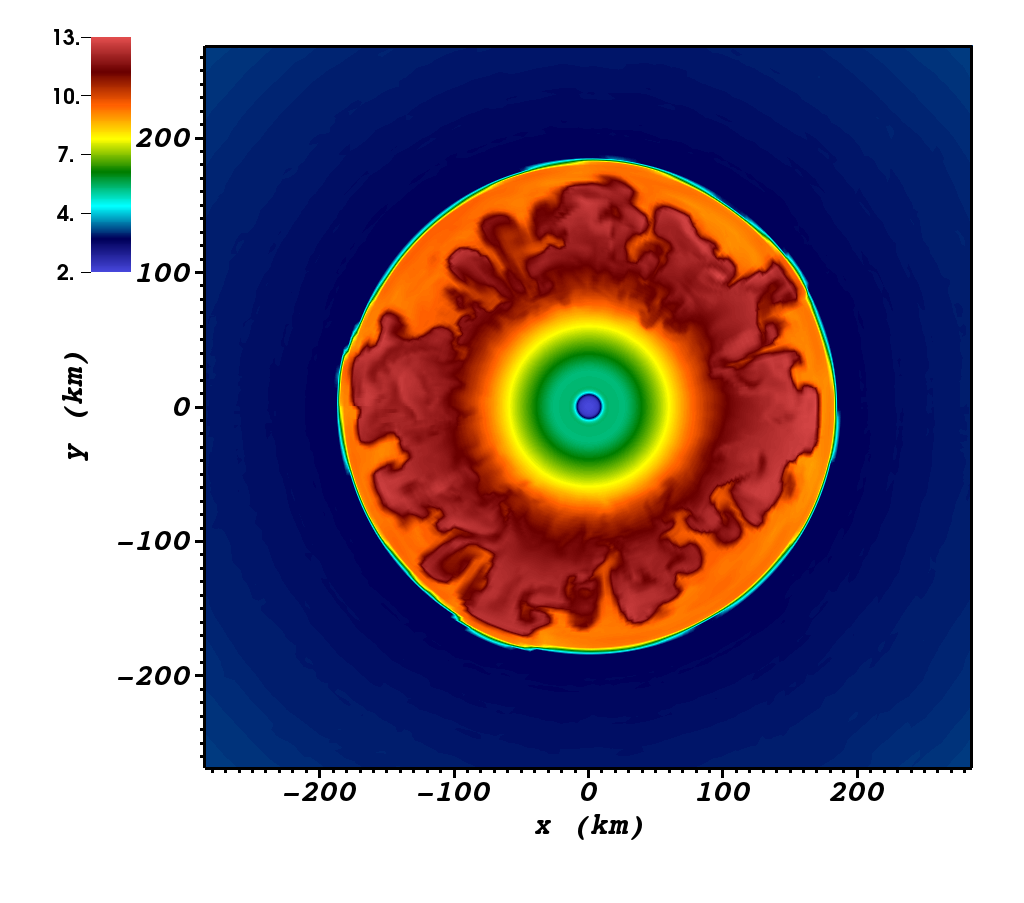}
\includegraphics[width=0.49\linewidth]{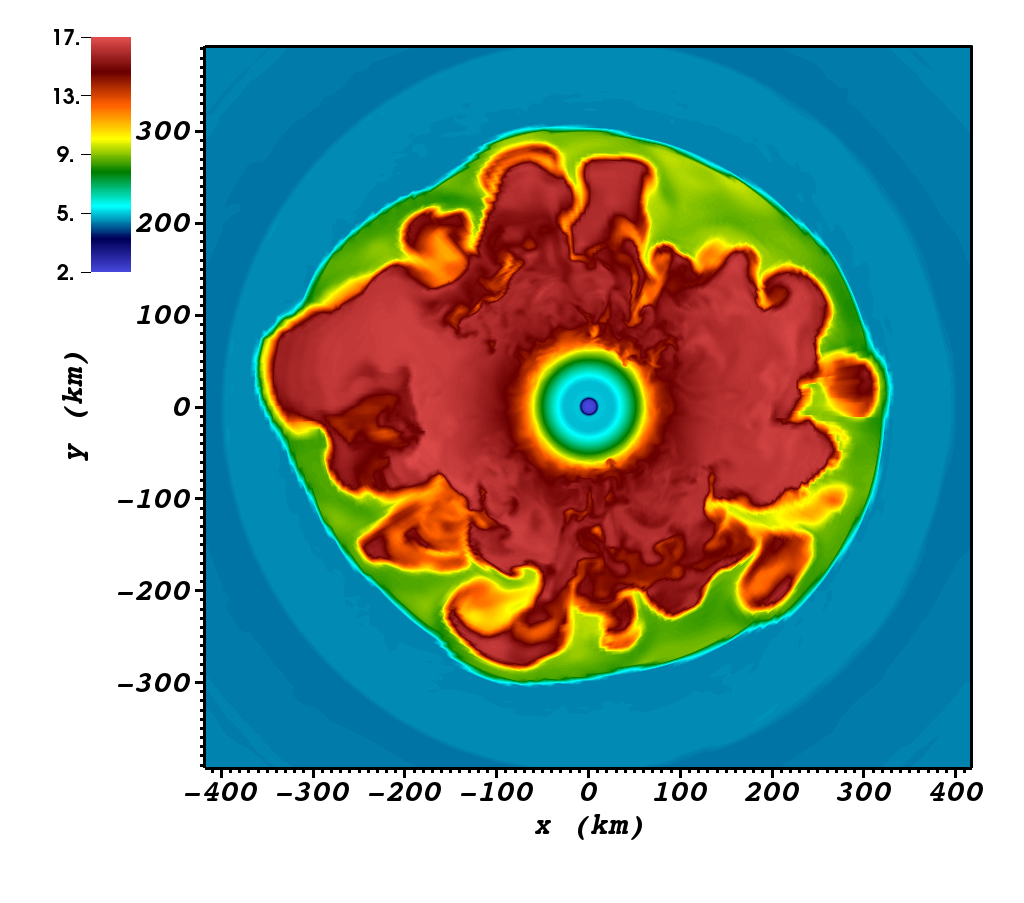}
\\
\includegraphics[width=0.49\linewidth]{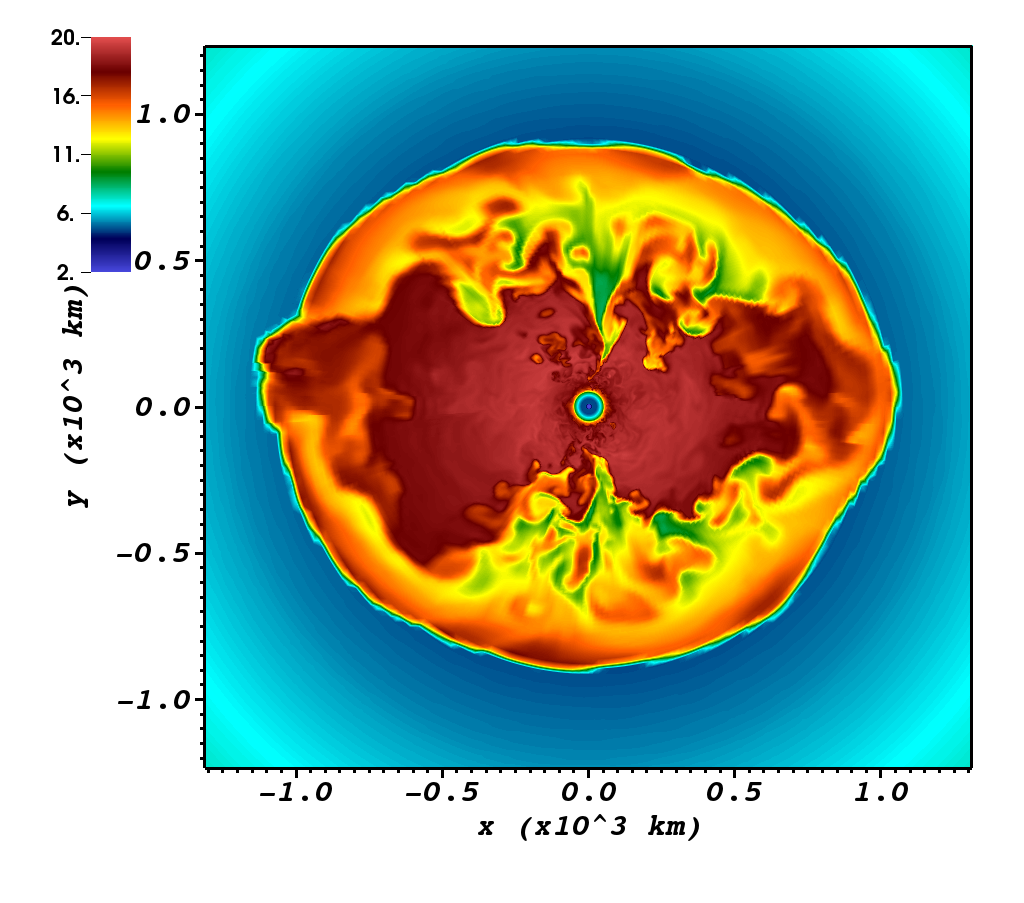}
\includegraphics[width=0.49\linewidth]{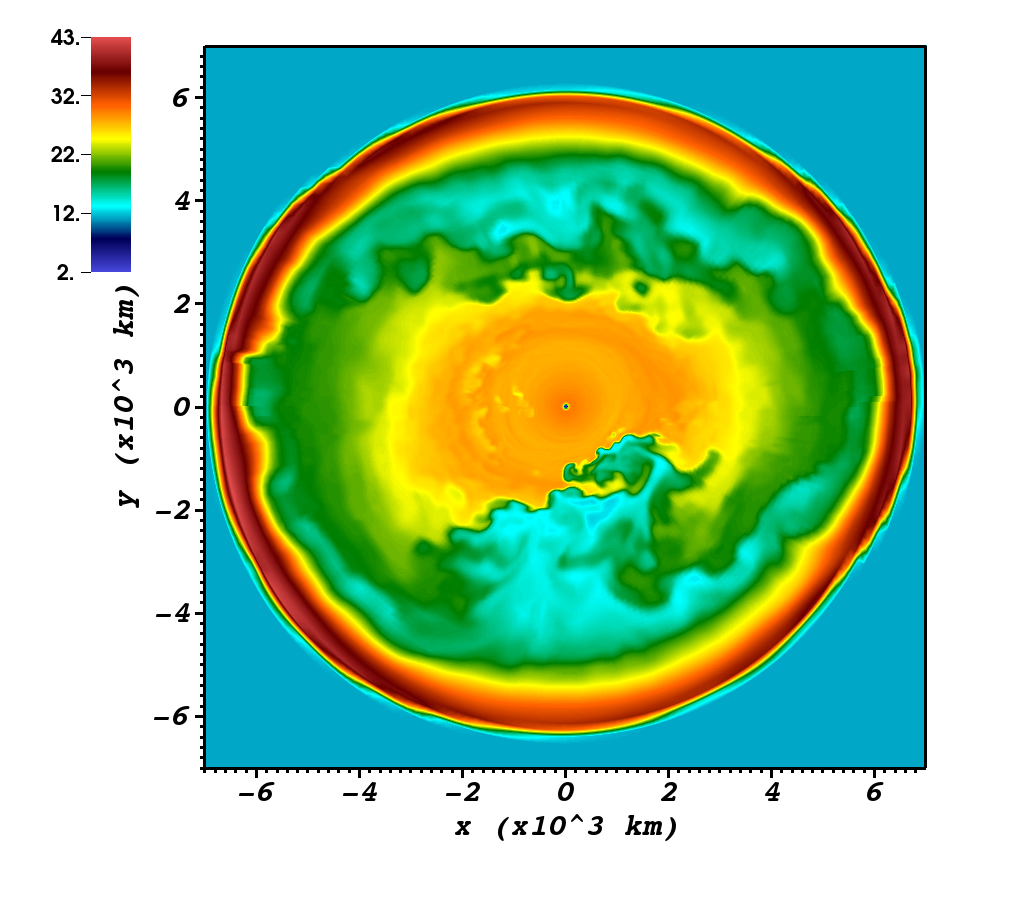}
\caption{2D slices showing the entropy in units
of $k_\mathrm{b}/\mathrm{nucleon}$ in model
s2.8-3D at post-bounce times
of $86 \, \mathrm{ms}$
(top left),
$133 \, \mathrm{ms}$ (top right),
$190 \, \mathrm{ms}$ (bottom left),
and $401 \, \mathrm{ms}$ (bottom right). The $x$-axis
is aligned with the axis of the spherical polar grid.
As the explosion develops and the shock radius grows,
a mild bipolar $\ell=2$ asymmetry in the flow develops.
At later times (bottom right), the bipolar deformation
becomes weaker, and the inner ejecta display a more
unipolar geometry with no apparent alignment with
the grid axis. At $401 \, \mathrm{ms}$, the 
transition to the neutrino-driven
wind phase is already underway.
\label{fig:snapshots}}
\end{figure*}

\section{Evolution During the First Second}
\label{sec:second}
\subsection{Explosion Dynamics}
The evolution of the diagnostic explosion energy (defined as in
\citealt{mueller_17}) 
for all models as well as the maximum, minimum and average
shock radii and trajectories of selected mass shells of the 3D model
are shown in Figure~\ref{fig:explosion}. For the 3D model,
we also present snapshots of the specific entropy on 2D slices
in Figure~\ref{fig:snapshots}.

As expected based on the pre-collapse density profile, the
ultra-stripped progenitor explodes in a manner very similar to the
bare C/O core models of \citet{suwa_15} and the low-mass single-star
progenitors just above the iron core formation limit
\citep{mueller_12b,melson_15a,mueller_16b,radice_17}. The shock moves outward
steadily, and shock expansion accelerates once neutrino-driven
convection develops about $80 \, \mathrm{ms}$ after bounce. Around
$150 \, \mathrm{ms}$, neutrino-heated material first reaches positive
net energy, and the diagnostic energy grows to $\mathord{\sim} 10^{50}
\, \mathrm{erg}$ by $300 \, \mathrm{ms}$. At this time accretion onto
the PNS has ceased, and the neutrino-driven wind has
developed and continues to pump some power into the explosion at a
modest rate.  The final explosion energy $E_\mathrm{expl}$ is
$1.12 \times 10^{50} \, \mathrm{erg}$, which is somewhat smaller than the value of
$1.77 \times 10^{50} \, \mathrm{erg}$ for model CO145 with a similar
C/O core mass of $1.45 M_\odot$, a difference which is likely to be
ascribed to the somewhat different neutrino transport treatment in our
model. Our final baryonic NS mass of $1.42 M_\odot$ is also
somewhat higher, reflecting considerable structural differences of our
progenitor compared to the bare CO-core models of \citet{suwa_15}. Using the fit formula for
the NS binding energy $E_\mathrm{bind}$ from
\citet{lattimer_01},
\begin{equation}
  E_\mathrm{bind} \approx 0.084 M_\odot c^2 (M_\mathrm{grav}/M_\odot)^2,
\end{equation}
this translates into a gravitational mass
of $M_\mathrm{grav}=1.28 M_\odot$, assuming
a final NS radius of $12\, \mathrm{km}$.
This value is in agreement with the typical mass measured for the young, second-formed NS in DNS systems \citep{tauris_17}.

We see a moderate increase of the explosion energy in 3D compared to
2D. This is in line with the small increase in explosion energy in 3D
found by \citet{melson_15a} due to the faster quenching of accretion
by 3D turbulence in the explosion of low-mass iron core progenitors
with fast shock propagation. The effect is somewhat larger than
in \citet{melson_15a}, especially for model s2.8-2D-a, which  also
shows the more unsteady growth of explosion energy due to
partial outflow quenching that is characteristic for more
massive progenitors \citep{mueller_15b}.

Compared to ECSNe
\citep{kitaura_06,fischer_10,huedepohl_10} and the structurally most
extreme low-mass iron core progenitors
\citep{mueller_13,melson_15a,mueller_16b,radice_17}, shock propagation is
slightly less rapid. At a post-bounce time of $150 \, \mathrm{ms}$,
the average shock radius is only around $400\, \mathrm{km}$ in
s2.8-3D compared to almost $1000 \, \mathrm{km}$ in the $9.6 M_\odot$
model of \citet{melson_15a}. As a result, there is sufficient time of
the 3D explosion model to develop a modest level of large-scale
asymmetries, resulting in a visible bipolar asymmetry at late times
(Figure~\ref{fig:snapshots}).  This is to be compared to the
small-scale asymmetries that dominate multi-dimensional ECSN models \citep{wanajo_11,wanajo_17} and the low-mass
iron core models of \citet{mueller_13,melson_15a,mueller_16b}.  Global
asymmetries are, however, less pronounced than in the ultra-stripped
models of \citet{suwa_15} with a ratio of the maximum and minimum
shock radius of no more than $\mathord{\sim}1.25$ around shock
revival.

\begin{figure}
\includegraphics[width=\linewidth]{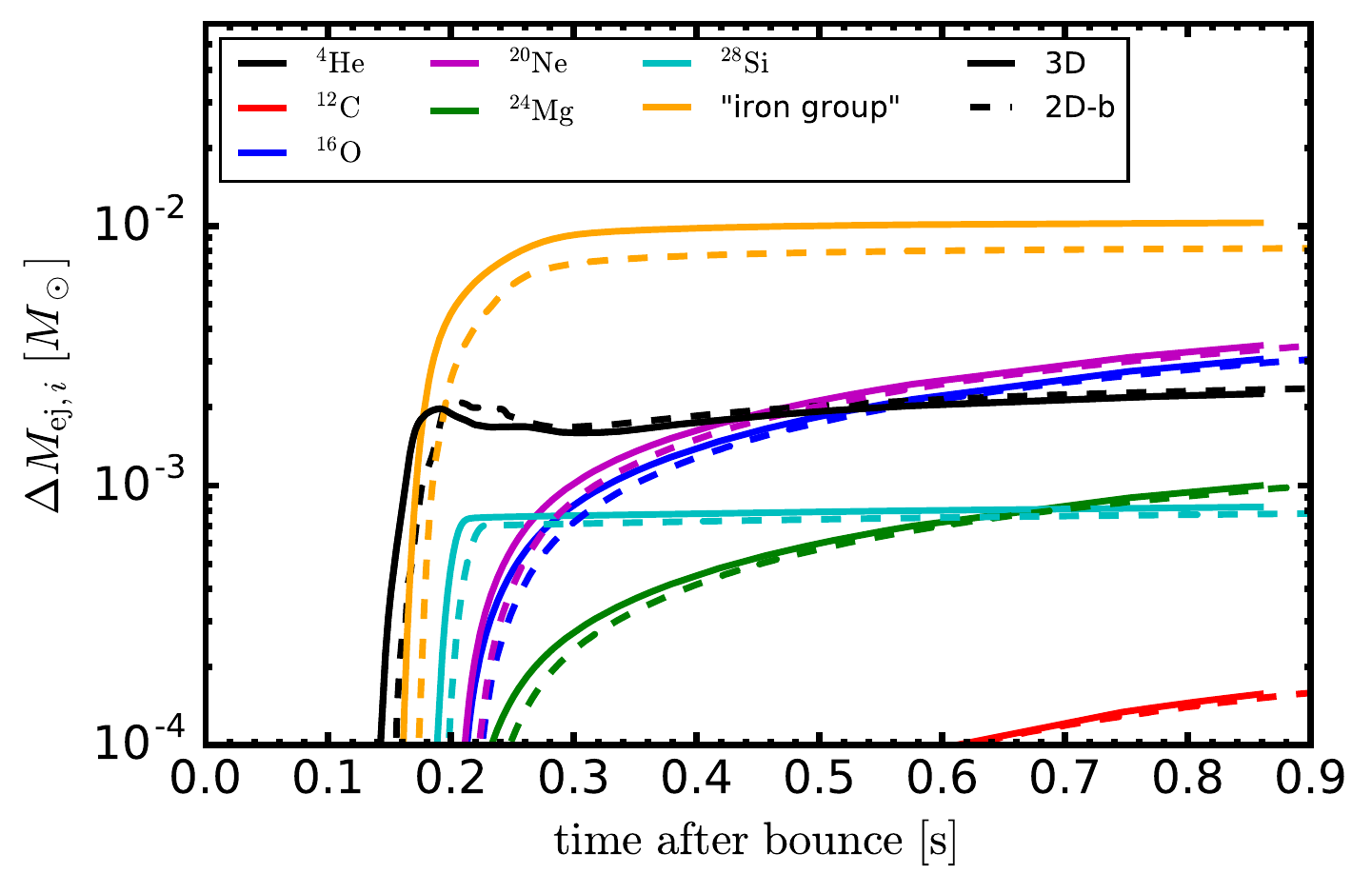}
\caption{Composition of the ejecta
(i.e., the formally unbound material behind the
shock) as a function of time for models
s2.8-3D (solid curves) and s2.8-2D-b (dashed). The curves of different
colour show the contribution 
$\Delta M_\mathrm{ej,i}$ of selected
species to the total ejecta mass.
The precise composition of the iron group ejecta
(orange) is not specified as it depends on the
electron fraction in the neutrino-heated ejecta,
which tends to be underestimated in the current
implementation of the \textsc{FMT} transport solver;
a sizeable fraction of the iron group ejecta
could be in the form of $^{56}\mathrm{Ni}$. 
A lower limit for the mass of $^{56}\mathrm{Ni}$
is provided by the amount of ejecta that undergo
explosive burning without being neutrino-processed,
which is roughly $10^{-3} M_\odot$.
\label{fig:composition}}
\end{figure}

The time-dependent composition of the ejecta is shown in
Figure~\ref{fig:composition}. The yields of iron group elements are
also similar to the simulations of \citet{suwa_15}. Roughly $10^{-2}
M_\odot$ of iron-group material is synthesised. Due to the simple
flashing treatment in our code and uncertainties in the electron
fraction due to our use of an approximate transport scheme, only the
total amount of iron group material can be given with some confidence,
and the detailed composition of this ejecta component remains
uncertain.  \citet{yoshida_17} found a significant amount of
neutron-rich ejecta and the production of the ligher trans-iron
elements similar to the case of electron-capture supernovae
\citep{wanajo_11} and low-mass iron core supernovae \citep{wanajo_17},
where the rapid expansion of Rayleigh-Taylor plumes after shock
revival leads to a freeze-out of the electron fraction below
$0.5$. Although our models exhibit similar explosion dynamics,
simulations with more accurate transport than our FMT scheme or the
IDSA approximation will be required to better assess the potential for
neutron-rich nucleosynthesis.
The amount of ${}^{56}\mathrm{Ni}$ produced in the explosion
 therefore remains uncertain as well. A lower limit
 for the nickel mass is provided by the amount of material
 with $Y_e=0.5$  that undergoes explosive burning to the
 iron group, which is about $10^{-3} M_\odot$.
 
Iron group nucleosynthesis has already finished by
the end of the 3D neutrino hydrodynamics simulation.
At this stage, only a few $10^{-3} M_\odot$ of intermediate-mass
elements have been swept up by the shock. During the
subsequent evolution the shells still outside the shock
will be completely ejected without any fallback.Including
the material ahead of the shock, the ejecta
will eventually comprise $0.024
M_\odot$ of helium, $0.011 M_\odot$ of oxygen, $0.01
M_\odot$ of neon, and smaller amount of
magnesium and carbon. Despite the small mass of the helium envelope,
helium thus remains the most abundant element in the ejecta.

\begin{figure}
\includegraphics[width=\linewidth]{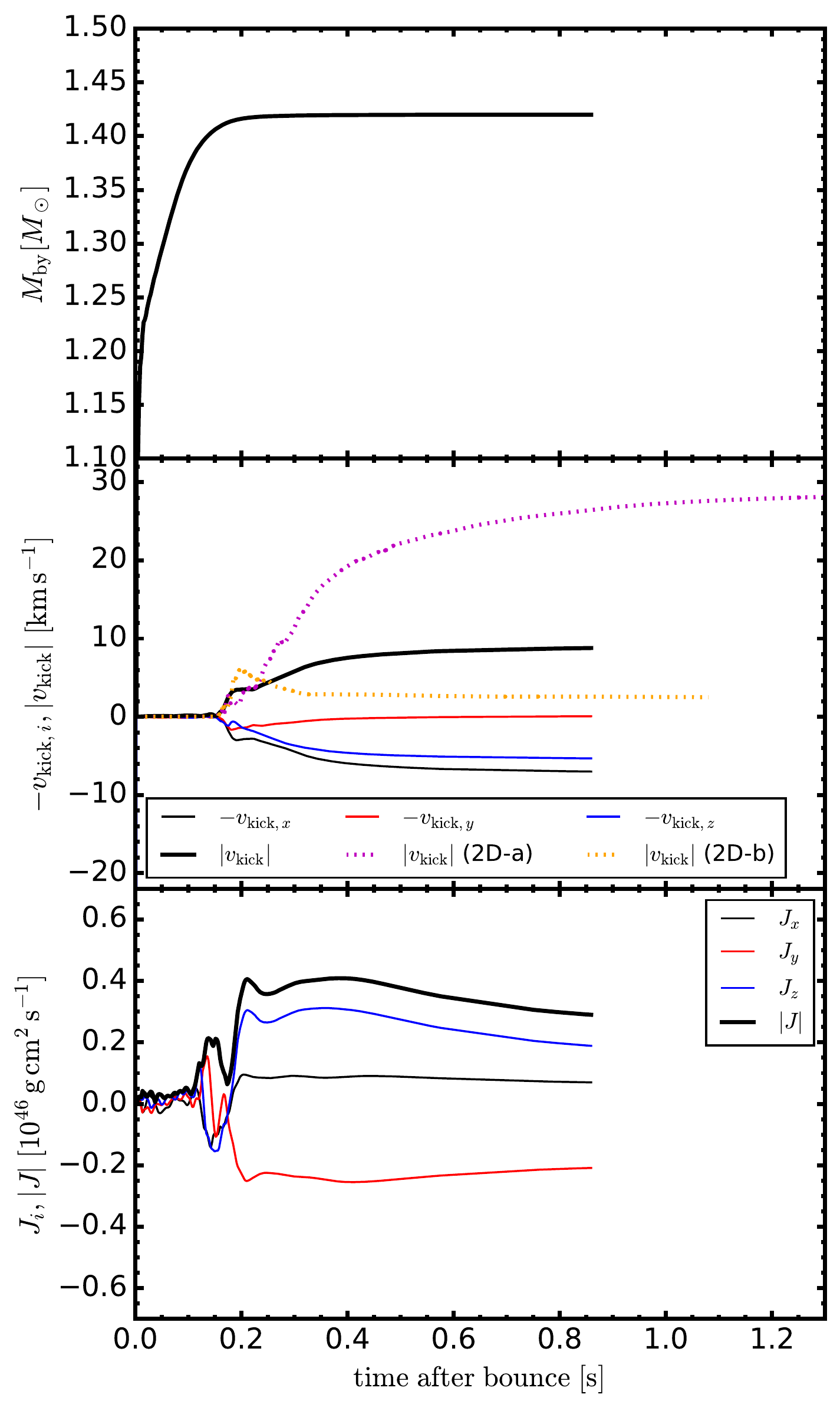}
\caption{Properties of the PNS
for the different simulations. Top panel: Baryonic PNS mass
$M_\mathrm{by}$ for model s2.8-3D; the results
for the 2D models are basically indistinguishable.
Middle panel: The components of the kick
velocity (thin lines) for model s2.8-3D and
the total kick velocity for s2.8-3D (thick black curve),
s2.8-2D-a (magenta, dotted) and
s2.8-2D-b (orange, dotted).
Bottom panel: The components (thin lines)
and absolute value (thick line) of the PNS
spin for model s2.8-3D according
to Equation~(\ref{eq:spin}).
\label{fig:ns}}
\end{figure}
\begin{figure}
\includegraphics[width=\linewidth]{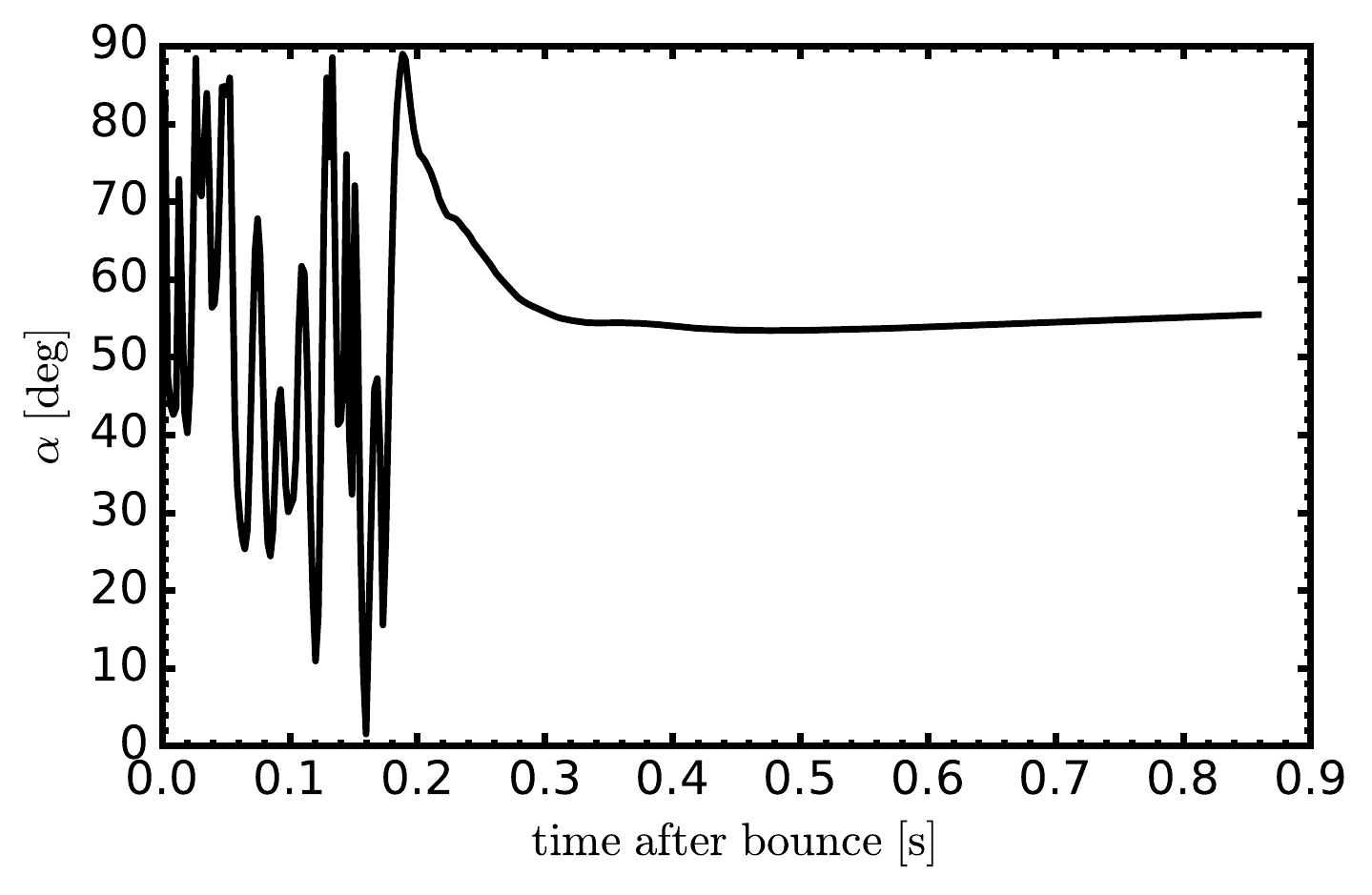}
\caption{Angle between the PNS angular momentum vector and kick velocity for model s2.83-3D. There is no tendency towards spin-kick alignment. 
\label{fig:alignment}}
\end{figure}

\subsection{Neutron Star Kick and Spin}\label{subsec:kick-spin}
Following \citet{scheck_06} and \citet{wongwathanarat_10b}, we evaluate the kick velocity $v_\mathrm{kick}$
of the PNS from the momentum of the ejecta
using momentum conservation,
\begin{equation}
  \label{eq:vkick}
  \mathbf{v}_\mathrm{kick}=-\frac{1}{M_\mathrm{PNS}}
  \left(
\int_\mathrm{ejecta} \rho \mathbf{v} \, \ud V+
\int\limits_0^t\oint \mathbf{F}_\nu \mathbf{n}\cdot \ud \mathbf{A} \, \ud t'
\right),
\end{equation}
where $M_\mathrm{PNS}$ is the (gravitational) PNS mass.
The momentum of the radiated neutrinos
is also included in the budget
in the second term in brackets,
where $\mathbf{F}_\nu$ denotes
the neutrino energy flux density
of all flavours on
a spherical shell far away
from the PNS (in our case
at $r=500 \, \mathrm{km}$). The
vector $\mathbf{n}$ denotes the unit
normal vector in the radial direction. Note
that this term contains a double
integral over the surface normal
vector $\ud \mathbf{A}$ and over
time $t'$.

For the 3D model, we can also calculate the
angular momentum $\mathbf{J}_\mathrm{PNS}$ of the PNS 
by integrating the angular momentum flux
through a sphere of radius $r_0$ around the
central remnant  \citep{wongwathanarat_10b,wongwathanarat_13},
\begin{equation}
\label{eq:spin}
  \frac{ \ud \mathbf{J}_\mathrm{PNS}}{\ud t}
  =
  \int \alpha \phi^4 r_0^2   \rho v_r \mathbf{v} \times \mathbf{r} \,\ud \Omega,
\end{equation}
where $\alpha$ and $\phi$ are
the lapse function and conformal
factor. The evolution of $\mathbf{v}_\mathrm{kick}$, $\mathbf{J}_\mathrm{PNS}$,
and the baryonic PNS mass $M_\mathrm{by}$ is shown
in Figure~\ref{fig:ns}. The kicks
have almost asymptoted to their final values
for all three models. Even extrapolating
the kick by assuming that the gravitational
acceleration $\mathbf{a}_\mathrm{PNS}$ of the PNS by the ejecta continues
after the end of the simulations and scales
with time $t$ as $t^{-2}$ (based
on  approximately homologous expansion of
the ejecta) does not change the values appreciably.

As expected from the general similarity of
the explosion dynamics with the C/O core models of \citet{suwa_15} 
of similar mass,
we also find small kick velocities with final values ranging
from $2.5$ to $28\, \mathrm{km} \, \mathrm{s}^{-1}$ for the s2.8-2D-b
and s2.8-2D-a models, respectively. Our 3D model results in a kick velocity of about
$9.4\,{\rm km\,s}^{-1}$. Since more models
would be needed to probe stochastic variations, these numbers are only
rough order-of-magnitude indicators for
the distribution of expected kick velocities.
It is already clear that the kicks
obtained for this particular progenitor
are somewhat larger than for the most
extreme single-star ECSN models in the literature
\citep{gessner_18}.
Different from ECSN models, the
contribution of anisotropic neutrino
emission to the kick is not negligible, though it remains by far
subdominant to the gravitational
tug on the PNS. In the 3D case, it amounts
to about $0.6\, \mathrm{km}\, \mathrm{s}^{-1}$.
The comparison with the
models of \citet{gessner_18} should not be
over-interpreted, however, since the progenitor
used in this study does not exhibit the
most extreme core-envelope structure among the
models of \citet{tauris_15}. Kicks
 of just a few $\mathrm{km} \, \mathrm{s}^{-1}$
may be generic for the ECSN channel of ultra-stripped progenitors, but this remains to be tested by future work.

Such small kicks for ultra-stripped supernovae 
of small iron cores are indeed in agreement
with 
earlier theoretical and observational arguments 
\citep{tauris_15,tauris_17} -- see also the 
proposed relation between NS mass and kick 
velocity for the second supernova in forming
DNS systems \citep{tauris_17}, as supported by 
current observational data of NS masses, 
eccentricities, proper motions and spin-orbit 
misalignment angles.

The 2D simulations of ultra-stripped supernovae of \citet{suwa_15} did
not allow any statement on the spin-up of the PNS due
to asymmetric accretion. Our 3D model shows that the spin-up is very
modest; the angular momentum imparted onto the PNS is
merely $0.27 \times 10^{46} \, \mathrm{g} \, \mathrm{cm}^{-2} \,
\mathrm{s}^{-1}$ (bottom panel of Figure~\ref{fig:ns}). Using the fit formula of \citet{lattimer_05} for the
moment of inertia $I$ of cold NSs,
\begin{equation}
  I\approx 0.237 M_\mathrm{grav} R^2 \left[1+4.2 \left(\frac{M_\mathrm{grav} \, \mathrm{km}}{M_\odot R}\right)
    +90\left(\frac{M_\mathrm{grav} \, \mathrm{km}}{M_\odot R}\right)^4\right],
\end{equation}
this translates into a spin period of $3 \, \mathrm{s}$ for
$M_\mathrm{grav}=1.28 M_\odot$ and assuming $R=12\, \mathrm{km}$. 
The
spin-up is thus considerably smaller than for massive progenitors with
sustained accretion \citep{mueller_17}.  It is also on the low side
compared to parameterised 3D explosion models of more massive
progenitors \citep{wongwathanarat_13}, which do not exhibit
considerable accretion after shock revival.  The small spin-up is a
natural result of the explosion dynamics of ultra-stripped models,
i.e., a fast explosion with small global asymmetries, the absence of
sustained accretion, and the small mass in the gain region that is
involved in overturn motions.

This result, 
however, comes with a large amount of uncertainty 
and is again only accurate to about an order of 
magnitude since we can neither explore
the stochasticity of the aspherical
accretion onto the PNS nor variations in the 
progenitor structure. Models with more massive
metal cores and stronger accretion after
shock revival could  presumably achieve
considerably shorter spin periods.
Moreover, the angular momentum
of the progenitor core will not be negligible in general.
For the shortest final orbital period of $0.035 \, \mathrm{d}$
considered in \citet{tauris_15}, the initial
angular momentum alone is sufficient to explain
NS spin periods as short as
$\mathord{\sim}400 \, \mathrm{ms}$.
Comparing to observations, there are two DNS systems 
known where the young NSs are detectable as radio 
pulsars and their spin periods are 144~ms and 
2.77~s. One must keep in mind that their spin 
periods at birth are smaller, potentially 
significantly smaller, depending on their true ages 
and the braking index.

From our simulations, we find no correlation between the direction of the spin axis and the kick velocity vector (Figure~\ref{fig:alignment}). This result is in agreement with recent simulations \citep{tauris_17} of the post-SN kinematics resulting from the second SN in DNS systems when calibrated to empirical data. A spin-kick alignment, however, has been suggested for isolated radio pulsars \citep[][and references therein]{noutsos_13}.

\begin{figure}
\includegraphics[width=\linewidth]{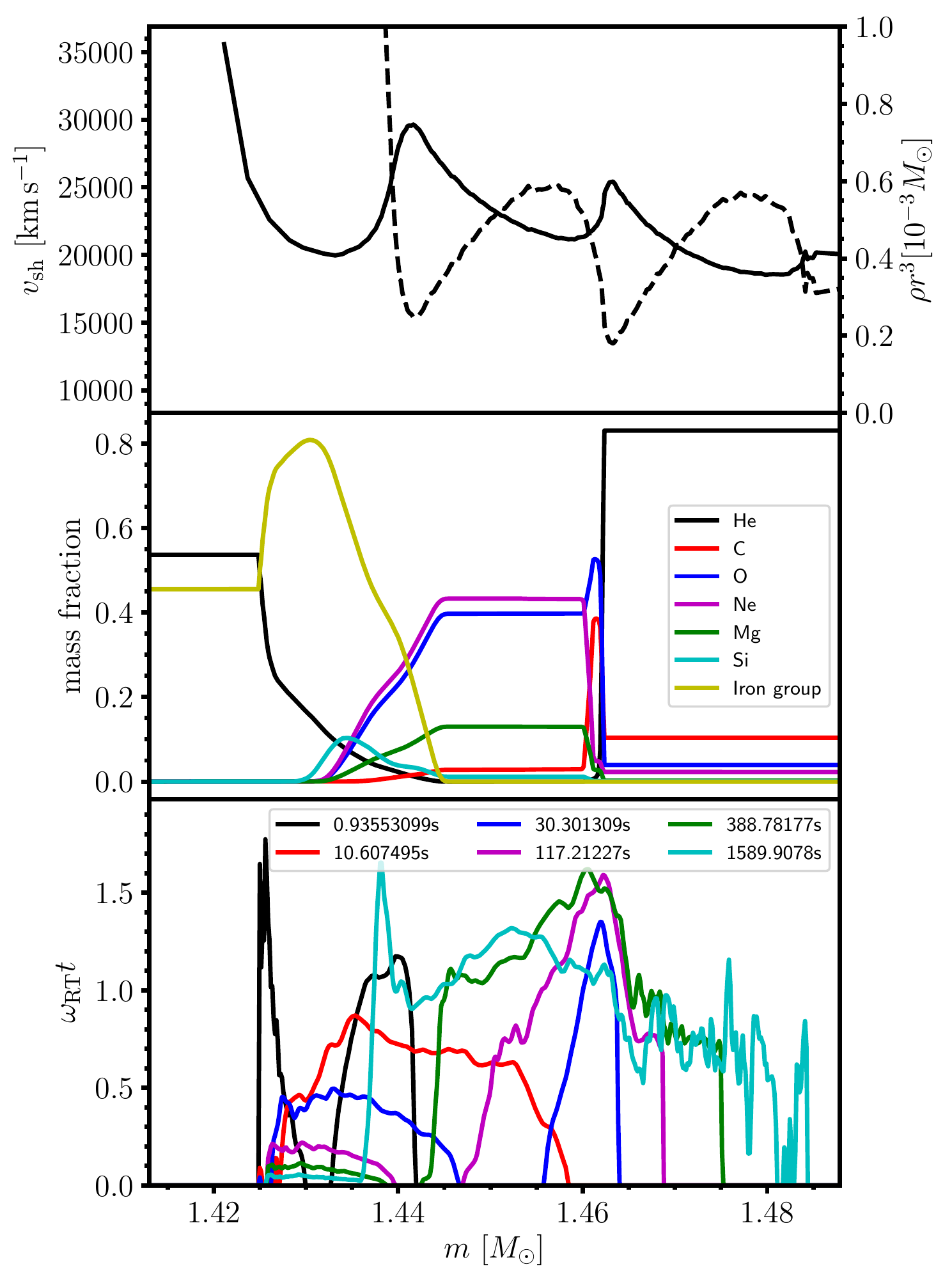}
\caption{Top panel:
Shock velocity (solid line) according to the fit formula of \citet{matzner_99}
(Equation~\ref{eq:vshock}) and variation of $\rho r^3$ in the progenitor model (dashed line).
Middle panel: Spherically averaged composition of model s2.8-3D
at the time of mapping into \textsc{Prometheus}  $730 \,\mathrm{ms}$
after bounce.
Bottom panel: Expected e-foldings $\omega_\mathrm{RT} t$ 
of the 
Rayleigh-Taylor instability per characteristic time-scale 
at various stages of the explosion
from the local growth rate $\omega_\mathrm{RT}$.
The local growth rate
$\omega_\mathrm{RT}$ is calculated from
spherically averaged profiles of the
3D model according to
Equation~(\ref{eq:omrt}). Rayleigh-Taylor instabilities
are triggered by the inner interface of
the O/Ne/Mg/C shell and the He envelope. The unstable
region below the He envelope extends inwards as the
reverse shock propagates deeper into the ejecta.
\label{fig:omrt}}
\end{figure}

\begin{figure}
\includegraphics[width=\linewidth]{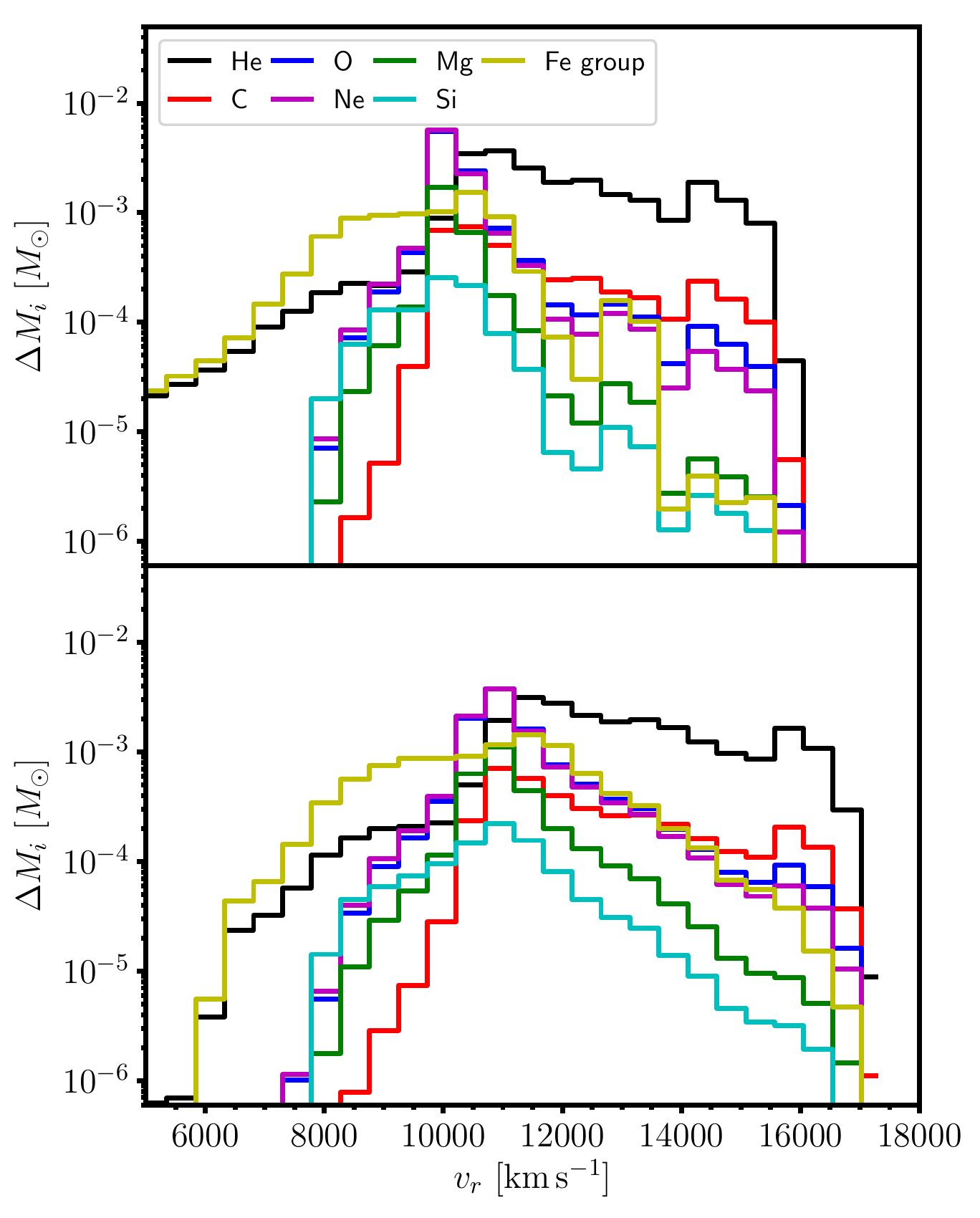}
\caption{Binned distribution of selected elements in the
ejecta as a function of radial velocity at the time of shock breakout 
in model s2.8-2D-b (top) and model S2.8-3D (bottom). 
\label{fig:mix_vel}}
\end{figure}

\begin{figure}
\includegraphics[width=\linewidth]{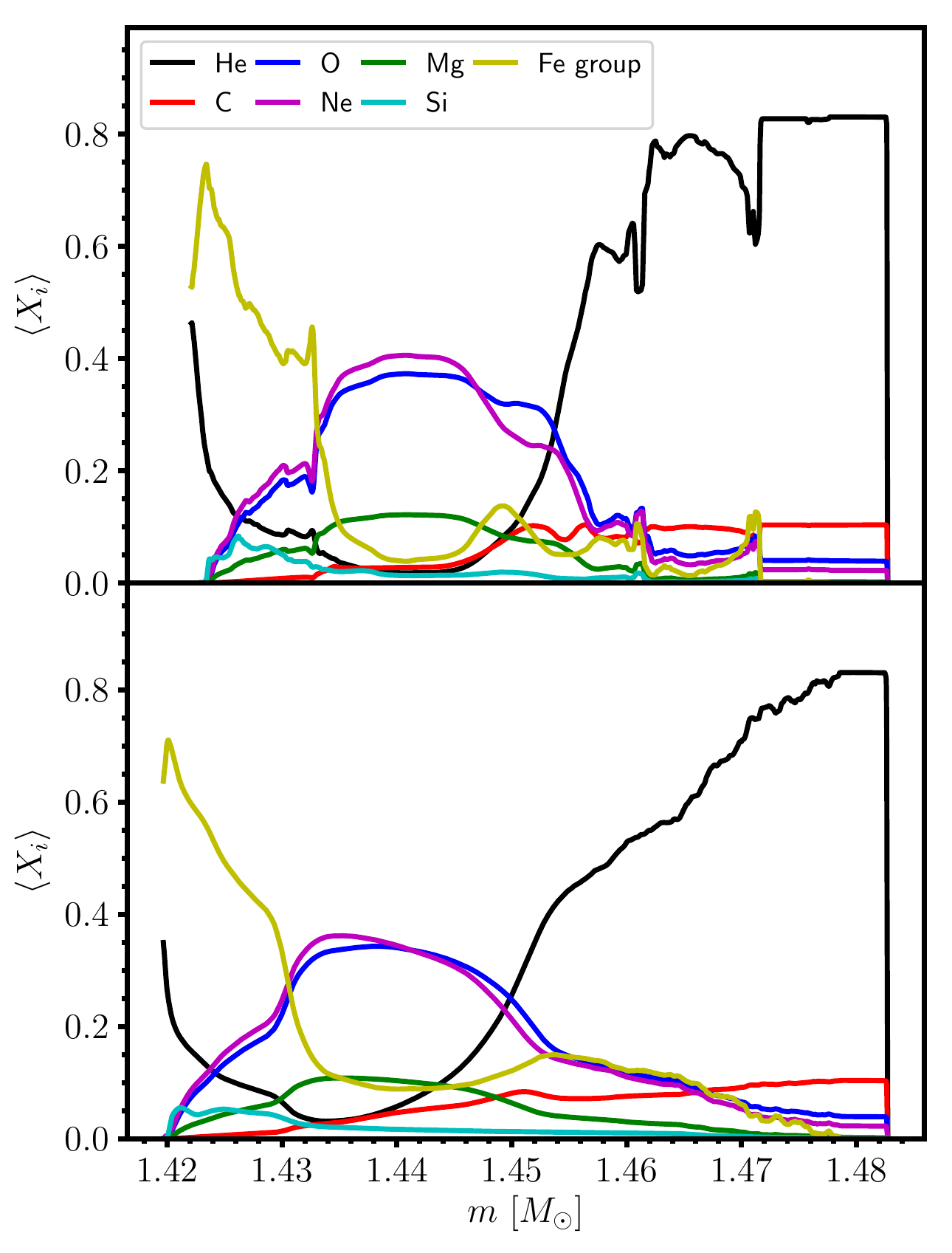}
\caption{Spherically averaged mass fractions of selected elements in the
ejecta at the time of shock breakout as a function of enclosed mass
in model s2.8-2D-b (top) and model S2.8-3D (bottom).
\label{fig:mix_mass}}
\end{figure}

\begin{figure}
\includegraphics[width=\linewidth]{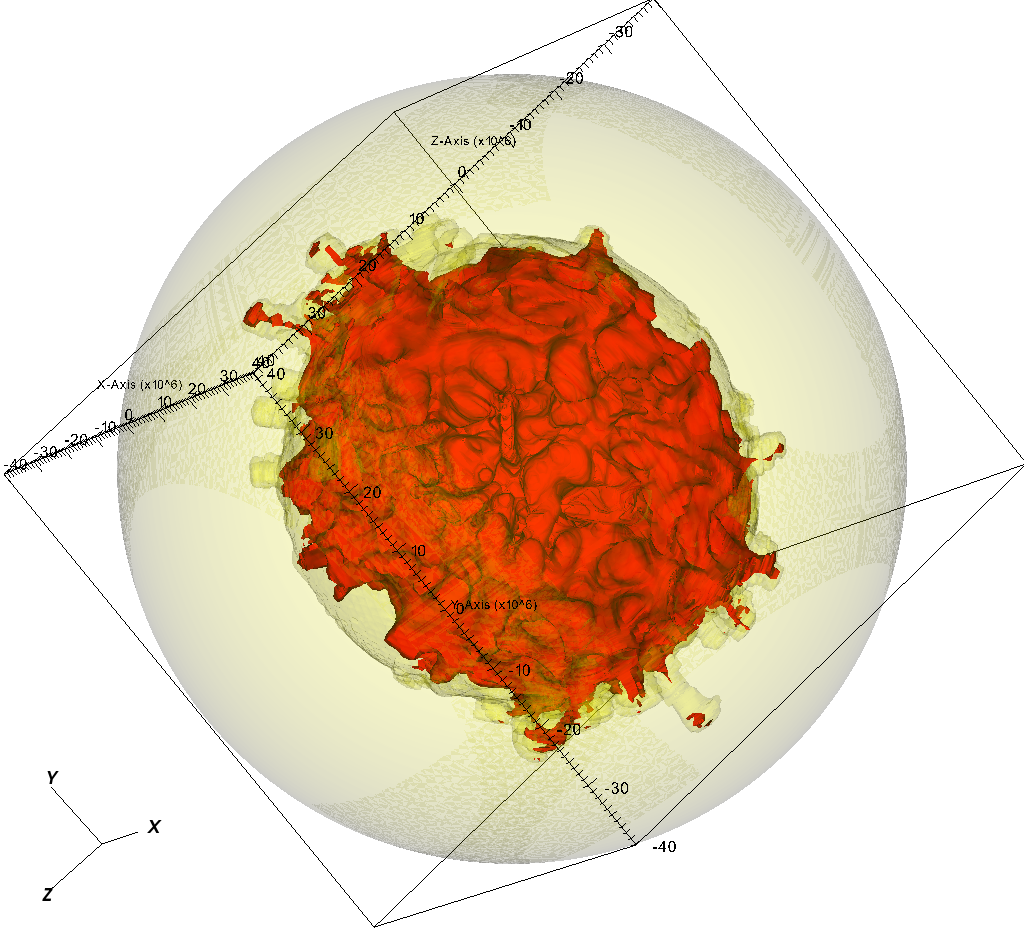}
\caption{Isosurfaces showing the mixing of iron-group ejecta at the
  time of shock breakout. The red isosurface denotes an iron group
  mass fraction of  $X_\mathrm{IG}=0.4$, The transparent
  yellow surfaces correspond to a helium mass fraction of
  $X_\mathrm{He}=0.5$, and the outer one coincides with the stellar
  surface. Only a few small-scale plumes of nickel/iron-rich ejecta
  have penetrated roughly half way through the helium envelope.
\label{fig:3d}}
\end{figure}

\begin{figure}
\includegraphics[width=\linewidth]{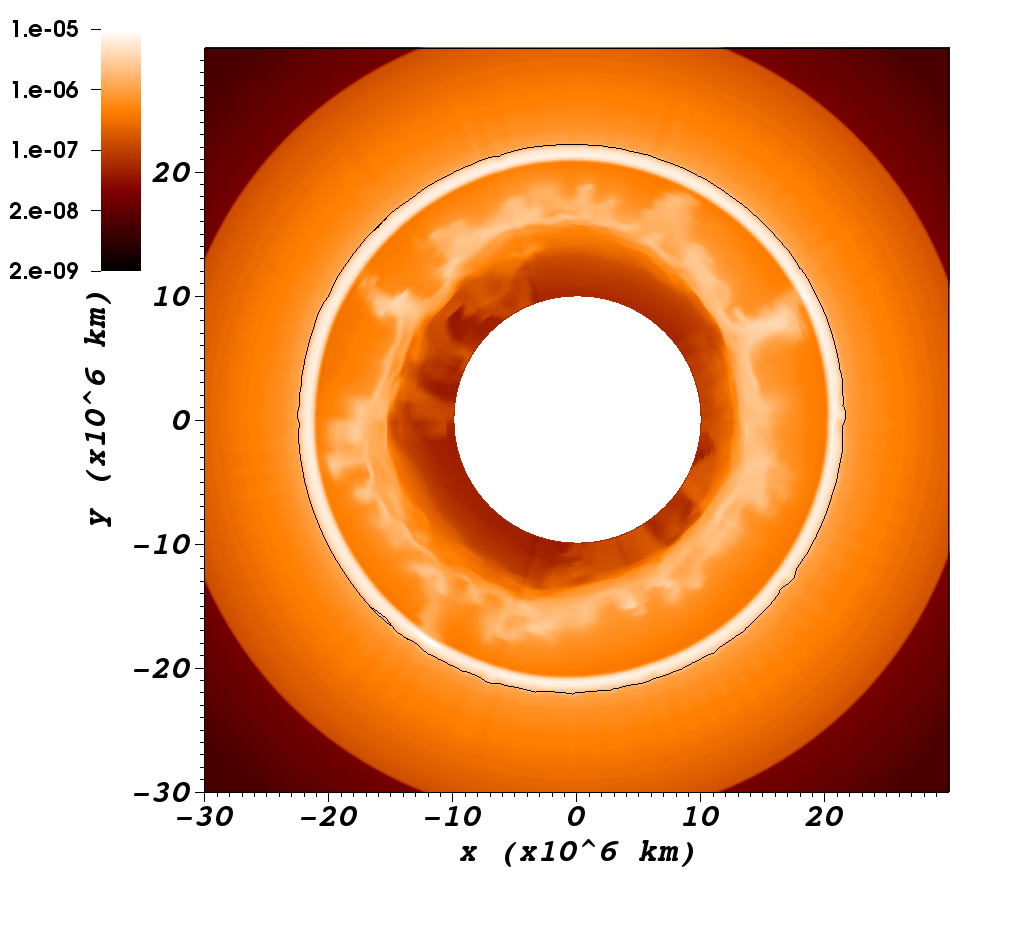}
\caption{2D slice showing the logarithm
($\log_{10}\rho$ in $\mathrm{g\,cm}^{-3}$) 
of the density $1700 \, \mathrm{s}$ after
the onset of the explosion. Most of the Rayleigh-Taylor plumes
formed at the inner interface of the O/Ne/Mg/C shell
are caught behind the dense shell (white annular
structure) formed by the
reverse shock at the core-envelope interface.
The black curve denotes the $X_\mathrm{He}=0.5$
isocontour for the mass fraction of He.
\label{fig:reverse_shock}}
\end{figure}

\section{Evolution to Shock Breakout}
\label{sec:envelope}
\subsection{Extent of Mixing}
As the shock propagates to the stellar surface, mixing driven by the
Rayleigh-Taylor instability occurs as the acceleration and subsequent
deceleration of the shock at shell interfaces establishes regions
where the pressure and density gradients point in opposite directions
\citep{chevalier_76,mueller_91,fryxell_91}. These episodes of
acceleration and deceleration are essentially determined by variations
in $\rho r^3$ in the density profile through which
the blast wave propagates \citep{sedov_59}, although the overall
decrease of the shock velocity $v_\mathrm{sh}$ is primarily due to the
accumulation of ejecta mass $M_\mathrm{ej}$. Both effects
can be well fitted by \citep{matzner_99},
\begin{equation}
\label{eq:vshock}
v_\mathrm{sh}
=
0.794
\left(\frac{E_\mathrm{expl}}{M_\mathrm{ej}}\right)^{1/2}
\left(\frac{M_\mathrm{ej}}{\rho r^3}\right)^{0.19}.
\end{equation}

For the ultra-strippped progenitor, this implies two episodes of shock
acceleration and deceleration (top panel of Figure~\ref{fig:omrt}).
The shock velocity peaks at $30,000 \, \mathrm{km} \, \mathrm{s}^{-1}$
at the base of the almost completely burned C shell and again
at $25,000 \, \mathrm{km} \, \mathrm{s}^{-1}$
at the base of the He envelope. Rayleigh-Taylor instabilities 
indeed occur in these regions as confirmed by local linear stability
analysis, which yields a growth rate of \citep{bandiera_84,benz_90,mueller_91}
\begin{equation}
\label{eq:omrt}
\omega_\mathrm{RT}=
\frac{c_\mathrm{s}}{\Gamma}
\sqrt{\left(\frac{\pd \ln P}{\pd r}\right)^2-\Gamma \frac{\pd \ln P}{\pd r}\frac{\pd \ln \rho}{\pd r}}
\end{equation}
for the compressible Rayleigh-Taylor instability, where $P$, $\Gamma$,
and $c_\mathrm{s}$ are the pressure, adiabatic index, and the speed of
sound. We evaluate Equation~(\ref{eq:omrt}) using spherically averaged
profiles of model s2.8-3D and show the expected number $\omega_{\mathrm RT} t$ of
e-foldings over one characteristic time-scale in the bottom panel of
Figure~\ref{fig:omrt}.  Rayleigh-Taylor
instabilities at the
two interfaces
develop already at $\mathord{\sim} 1\, \mathrm{s}$ and
$\mathord{\sim} 30\, \mathrm{s}$,
respectively.

Especially at the inner edge of the O/Ne/Mg/C shell, the nominal growth
factors appear small (about 2.5 e-foldings during the first $400 \,
\mathrm{s}$), but this is merely due to the fact that the instability
quickly becomes non-linear due to the asphericities seeded by the
neutrino-driven engine, so that the feedback of mixing on the density
and pressure profiles reduces the nominal linear growth rate.\footnote{Note that many papers in the literature
  \citep[e.g.\ ][]{benz_90,mueller_91,fryxell_91,kifonidis_03,wongwathanarat_15}
  provide growth rates based on 1D models with similar energetics as
  their multi-D simulations and therefore obtain a larger number of
  e-foldings.}  There is in fact considerable mixing of iron group
elements into and through the O/Ne/Mg/C shell as can be seen from the
final distribution of the ejecta at shock breakout in velocity space
(Figure~\ref{fig:mix_vel}) and as a function of mass coordinate
(Figure~\ref{fig:mix_mass}).  An appreciable amount of O, Ne, Mg, and
of Fe group material makes it far into the He envelope, with somewhat
more efficient mixing in the 3D model. Merely judging by the mass
fractions of iron group elements of $\mathord{\sim} 0.1$ in the He
shell, the mixing of ${}^{56}\mathrm{Ni}$ appears more than sufficient
to make the He visible in the spectra, which requires mass fractions
$\gtrsim 0.01$ \citep{dessart_12,dessart_15}. 
The 3D distribution of
the mixed Fe group material precludes any firm conclusions on the
impact of the mixing on the spectra at this stage, however: The
$^{56}\mathrm{Ni}$ that is mixed into the He shell is concentrated
into a few thin plumes (Figure~\ref{fig:3d}) with significantly higher
density than the ambient He, and it still needs to be determined
whether such a strongly clumped distribution of $^{56}\mathrm{Ni}$ can
lead to efficient non-thermal excitation of He in a large fraction of
the envelope.

The extent of the mixing falls between the few studies that have
addressed stripped-envelope supernovae of Type Ib/c and IIb with
consistent \citep{ellinger_13} or artificially altered
\citep{kifonidis_03,wongwathanarat_17} envelope structures. Whereas --
at the extreme end -- \citet{ellinger_13} found little mixing during
the explosion in a Type~Ib supernova model due to the lack of a maxing
episode at an H/He interface, ${}^{56}\mathrm{Ni}$ is thoroughly
homogenised from the mass cut far into the He shell in a Ib supernova
model of \citet{kifonidis_03} with an artificially truncated envelope
(see their figure~19). Our model is less extreme than that of
\citet{kifonidis_03}, as one still recognises three fairly distinct
layers, i.e., the He shell with C enrichment from the active shell
source, the remains of the O/Ne/Mg/C shell, and the inner ejecta that
mainly consist of $^{56}\mathrm{Ni}$ from explosive burning, Fe group
material and He from neutrino-driven outflows, and O, Ne, Mg, and Si
swept up by the shock at early times. More efficient mixing into the
He shell is impeded by the development of the reverse shock from the
base of the He envelope, which confines all but the fastest Ni-rich
plumes within the O/Ne/Mg/C shell, similar to the situation at the
He/H shell interface of some blue supergiant models
\citep{kifonidis_03,wongwathanarat_15}. This phenomenon illustrates
that is is crucial that mass loss is included appropriately in the
evolution of the progenitor and that the structure of the He envelope
in stripped-envelope supernovae is modelled consistently. The
formation of the reverse shock depends critically on the relatively
strong acceleration and deceleration of the shock at the base of the
He envelope, which is in turn tied to the expansion of the envelope to
a radius of $5\times 10^{12} \, \mathrm{cm}$, which could not be
obtained by cutting the hydrogen envelope of an ``appropriate''
hydrogen-rich single-star progenitor model.

Such a variety of outcomes in stripped-envelope models is not
unexpected, and it would be premature to make general statements about
the effects of envelope stripping on mixing instabilities. There is no
reason to expect considerably less variation in mixing than between
hydrogen-rich red and blue supergiant progenitors: The development of
the Rayleigh-Taylor instability at shell interfaces inside the
C/O-core is bound to be very sensitive to variations in the structure
and configuration of the interior burning shells (which can be
considerable, see \citealt{collins_18,sukhbold_17}) and the seed
asphericities imprinted by the supernova engine. One also expects
that the character of the core-envelope interface -- and hence
of the mixing and the reverse shock associated with it -- varies
considerably since the radial extent of the He envelope of
Type~Ib/c supernova progenitor should span about two orders of
magnitude \citep{yoon_10}.

\begin{figure}
\includegraphics[width=\linewidth]{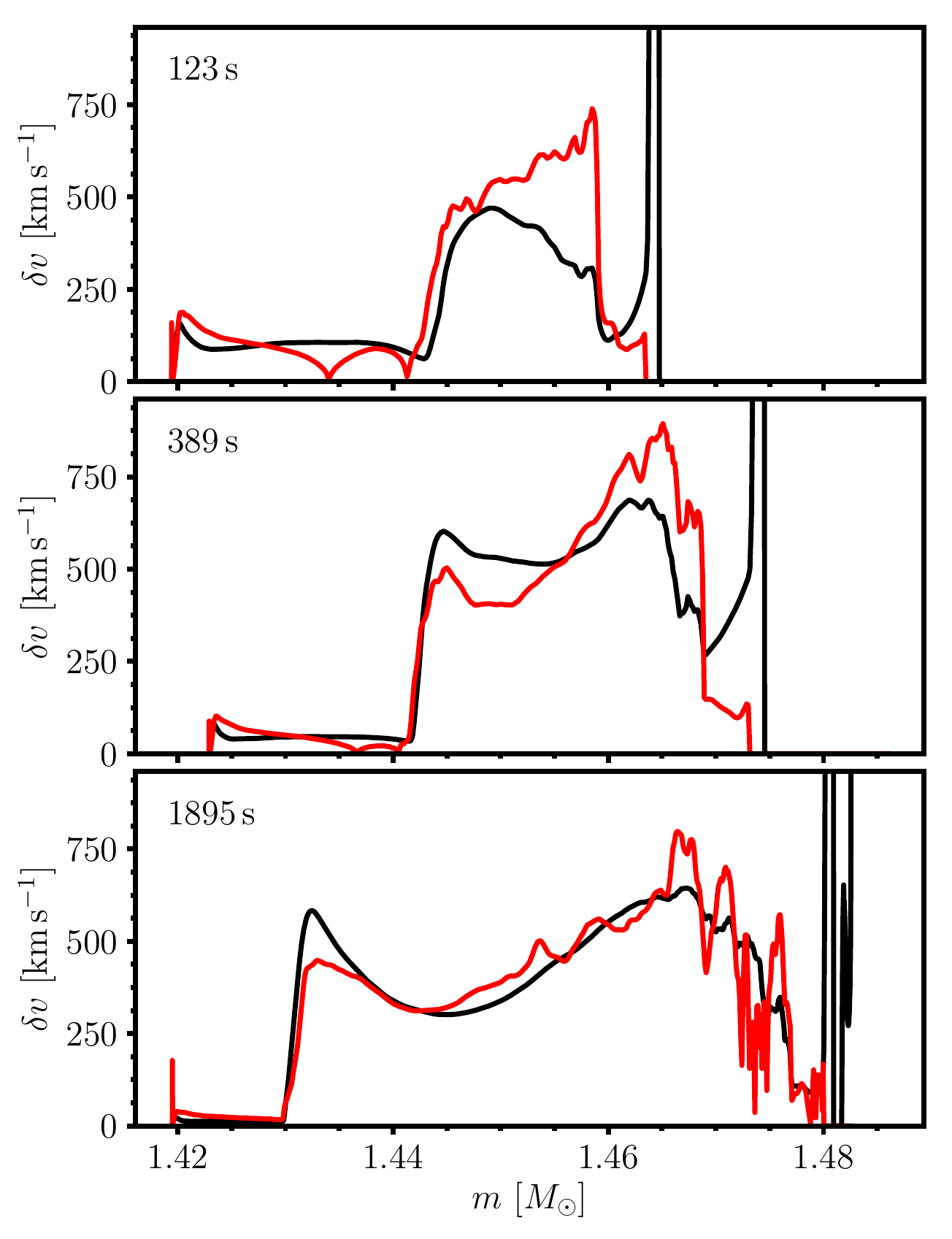}
\caption{Comparison of the
root-mean-square fluctuations $\delta v$
of the radial velocity from model
s2.8-3D (black) to the estimate from 
Equation~(\ref{eq:vbal}) based on a simple
buoyancy-drag-model (red) at three selected
times.
\label{fig:balance}}
\end{figure}

\subsection{Non-Linear Regime of the Rayleigh-Taylor Instability and Assessment of 1D
  Mixing Models} 

\citet{paxton_17} recently suggested that mixing by the
Rayleigh-Taylor instability is amenable to a 1D treatment based on an
appropriate turbulence model \citep{duffell_16}. The possibility of an
effective 1D treatment has largely remained unexplored during the
long history of multi-D simulations of mixing instabilities in
core-collapse supernovae. Although \citet{paxton_17} presented an
encouraging comparison of their 1D mixing algorithm in the
\textsc{Mesa} code with 3D models of \citet{wongwathanarat_15}, it is
imperative to further investigate the validity and robustness of such
an effective 1D treatment.  To this end, we shall take a closer look
at the turbulent velocities and turbulent fluxes in our 3D
model. Instead of merely comparing to the final result of the algorithm in
\textsc{Mesa}, we rather examine the individual physical assumptions
of the underlying turbulence model to complement the discussion in
\citet{paxton_17}.

\begin{figure}
\includegraphics[width=\linewidth]{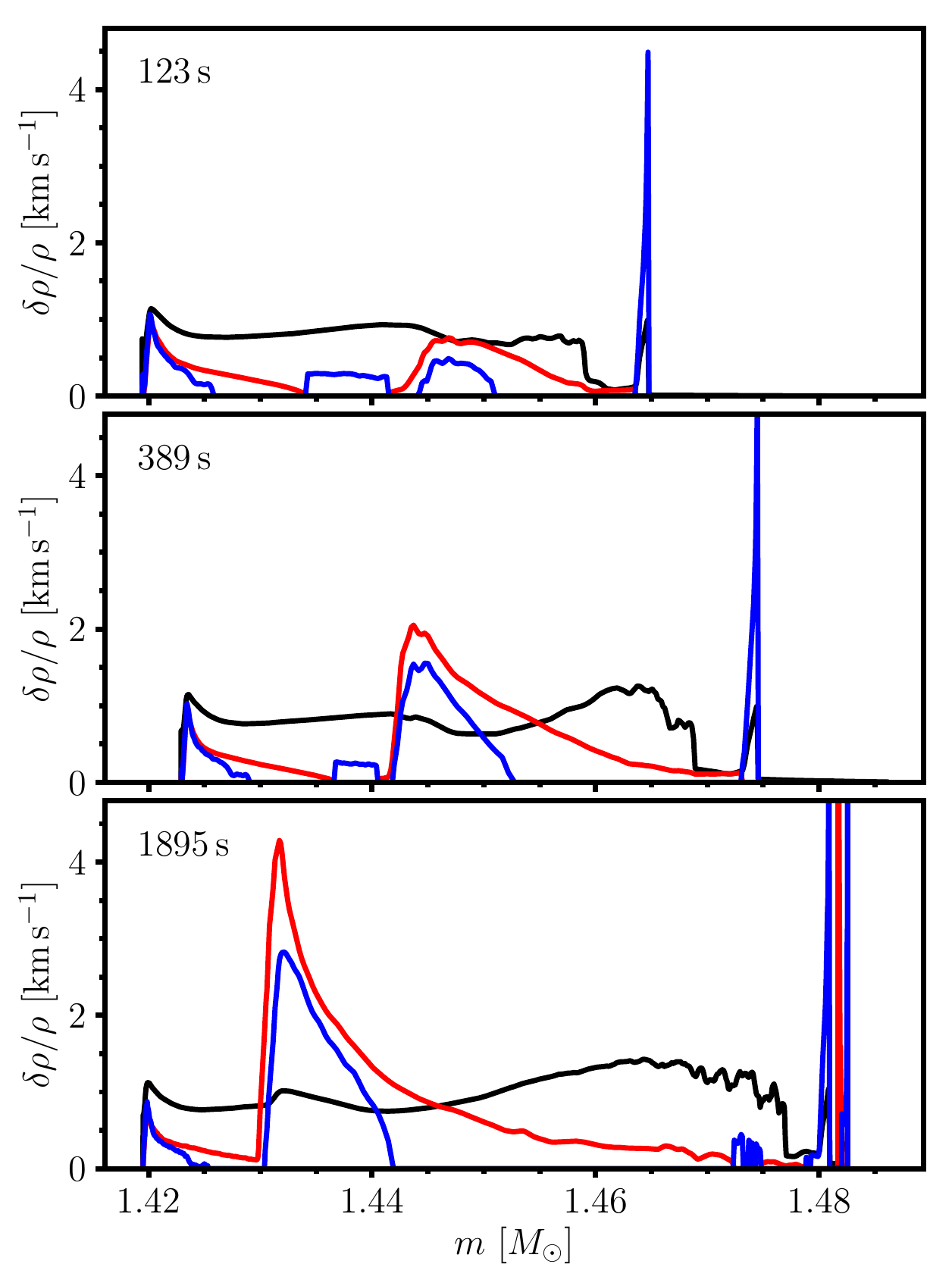}
\caption{Comparison of the relative
root-mean square density contrast $\delta \rho/\rho$
in model s2.8-3D (black) to
mixing-length approximations
based on the local gradients of the
spherically averaged density and pressure.
The red and blue curves show
the results for the equations
including the adiabatic expansion
of bubbles
(Equation~\ref{eq:drho_c})
and the estimate from \citet{duffell_16}
(Equation~\ref{eq:drho_ic}), respectively.
In formally stable regions
where these stability criteria give $\delta \rho/\rho<0$,
the value is set to zero instead.
\label{fig:dflukt}}
\end{figure}

\begin{figure*}
\includegraphics[width=0.49\linewidth]{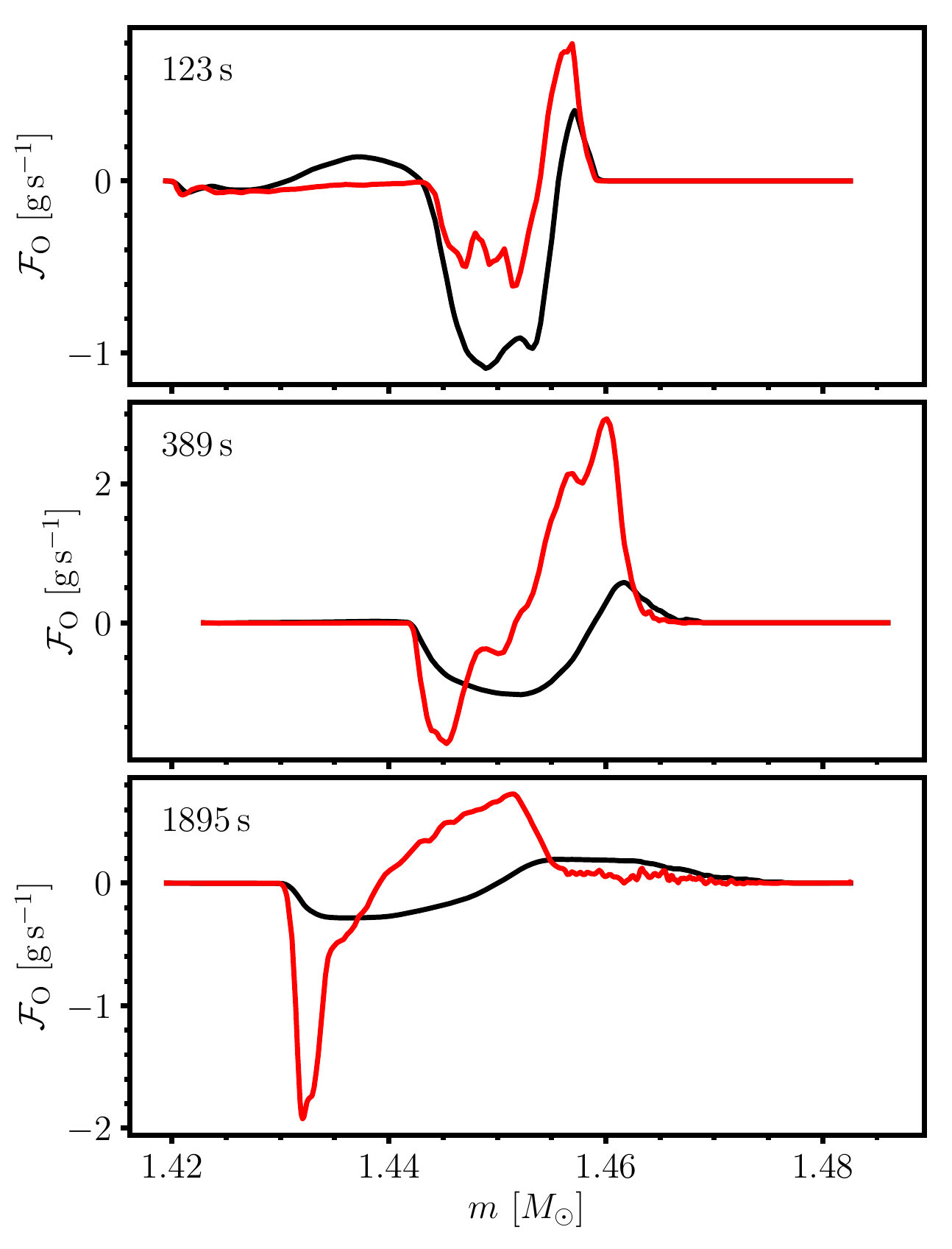}
\includegraphics[width=0.49\linewidth]{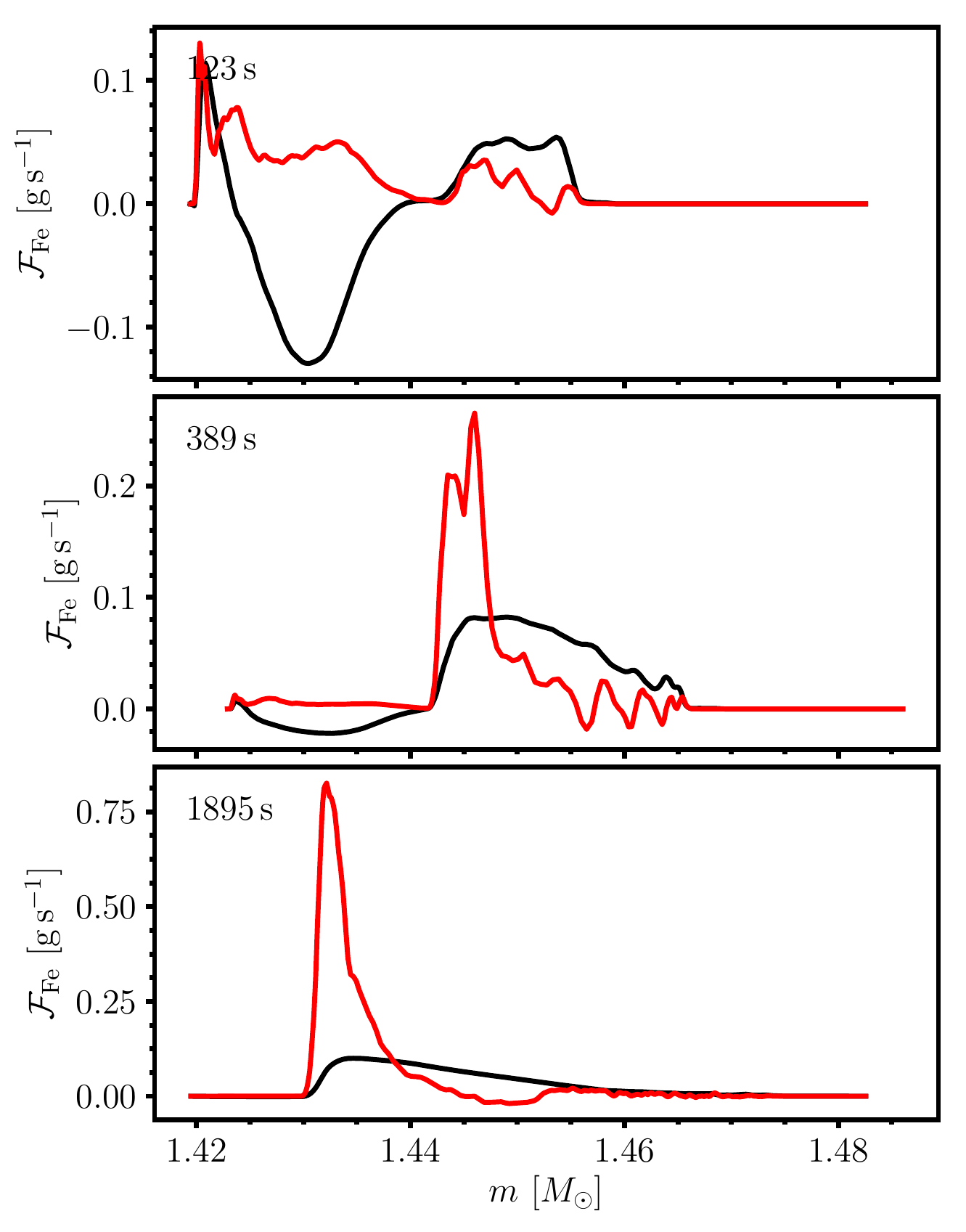}
\caption{Turbulent mass fluxes $\mathcal{F}_\mathrm{O}$ (left)
and $\mathcal{F}_\mathrm{Fe}$ (right) for ${}^{16}\mathrm{O}$
and ${}^{60}\mathrm{Fe}$ (a tracer for
the neutrino-processed iron/nickel-rich ejecta)
in model s2.8-3D (black) compared to
the generalised mixing-length approximation (red)
from Equation~(\ref{eq:fmix}) at selected times.
\label{fig:xflux}}
\end{figure*}

The approach of \citet{paxton_17} and \citet{duffell_16} ultimately
lumps the effects of the mixing instabilities into  diffusion terms.
The diffusive flux $F_Y$ for quantity $Y$ is expressed in terms
of the turbulent fluctuations $\delta v$ of the (radial) velocity and
the contrast $\delta Y$ between the plumes that are mixed outward
and inward; and in the vein of mixing-length theory, $\delta Y$
is estimated from the local gradient and a mixing length $\Lambda$,
\begin{equation}
\label{eq:fdiff}
F_Y=\delta v \, \delta Y=\delta v \Lambda \frac{\pd Y }{\pd r}.
\end{equation}
The mixing length is chosen as
\begin{equation}
\label{eq:lambda}
\Lambda =C \frac{\delta v}{c_\mathrm{s}} r,
\end{equation}
where $C$ is an appropriate non-dimensional coefficient. This choice
is motivated by the realisation that the mixing length should depend
on the distance that Rayleigh-Taylor plumes can traverse within
one characteristic time-scale of the system.

For the turbulent velocity fluctuations, \citet{paxton_17}
solve a time-dependent equation for the ratio
$\kappa=\delta v^2/c_\mathrm{s}^2$,
\begin{eqnarray}
\label{eq:vconv}
&&\frac{\pd \rho \kappa}{\pd t}
+
\frac{1}{r^2}\frac{\pd}{\pd r}
\left[
r^2(\rho \kappa v_r-C \kappa c_\mathrm{s} r \frac{\pd \rho \kappa}{\pd r})
\right ]
\\
\nonumber
&&
\quad =
(A+B \kappa)\sqrt{\max (0,-\frac{\pd P}{\pd r}\frac{\pd \rho}{\pd r})}
-D \kappa \rho c_\mathrm{s} r^{-1}.
\end{eqnarray}
where $A$, $B$, and $D$ are again appropriately chosen non-dimensional
coefficients. \citet{paxton_17} set those non-dimensional coeffiecients to
$A=10^{-3}$, $B=2.5$, $C=0.2$, $D=2$.

Whereas the validity of Equation~(\ref{eq:fdiff}) can be tested
directly by comparing to the actual turbulent fluxes in multi-D
simulations (see below), it is less straightforward to pit
Equation~(\ref{eq:vconv}) against a multi-D model without actually
solving the time-dependent equations of the turbulence model.  In
fact, our simulations show that the turbulent velocity fluctuations
appear to be captured by an even simpler model in the non-linear
regime of the Rayleigh-Taylor instability. Judging by the morphology
of the plumes in typical supernova simulations, the Rayleigh-Taylor
instability typically proceeds far into the second stage,
where the plume velocity is determined by the balance between
buoyancy and the drag force \citep{sharp_84,zhou_17a,zhou_17b}.  This
implies
\begin{equation}
\delta v = \sqrt{\lambda \frac{\delta \rho}{\rho} g_\mathrm{eff}},
\end{equation}
in terms of the density contrast $\delta \rho$, the spherically averaged
density $\rho$, the effective acceleration $g_\mathrm{eff}=
\rho^{-1} \pd P/\pd \rho$, and
a length scale  $\lambda$ that encapsulates the drag coefficient and the
volume-to-surface ratio of the plumes. Expressing $g_\mathrm{eff}$
in terms of $\rho$ and the gradient of the spherically averaged pressure $P$,
this becomes
\begin{equation}
\label{eq:vbal}
\delta v = \sqrt{\lambda \frac{\delta \rho}{\rho} \left |\frac{1}{\rho}\frac{\pd P}{\pd
    r}\right| }.
\end{equation}
In Figure~\ref{fig:balance} we evaluate Equation~(\ref{eq:vbal})
for three representative times using 
root-mean-square fluctuations (RMS) from a spherical
Favre decomposition for $\delta \rho$ and $\delta v$.
It can be seen that this simple buoyancy-drag model quite
accurately captures the evolution of $\delta v$  for $\lambda = 0.016 r$. The small value of
$\lambda$ is consistent with the morphology of Rayleigh-Taylor mixing
in our model, which is dominated by small-scale plumes. These findings
suggests that balance between buoyancy and drag is indeed what
determines the evolution of the Rayleigh-Taylor plumes in this particular
model.

But what does this finding imply for effective 1D turbulence
models for Rayleigh-Taylor mixing and can it be related to
Equation~(\ref{eq:vconv})? The concept of a balance between
effective buoyancy and drag is not included in 
Equation~(\ref{eq:vconv}) by construction; indeed one
notices that the dominant source and sink terms on the right-hand side
cannot balance each other because they are both proportional
to $\kappa$ (and the small term proportional to $A$ merely serves
to kick off the growth of the instability and is not important in
the non-linear regime). The only terms that can lead to a balance
condition for the plume velocity are the linear source term
and the diffusive term, which is quadratic in $\kappa$. If
we match these two terms using dimensional analysis (i.e., replacing
radial derivatives with $r^{-1}$), we arrive at the condition
\begin{equation}
C c_\mathrm{s} r^{-1} \rho \kappa^2
\sim
B \kappa \sqrt{-\frac{\pd P}{\pd r}\frac{\pd \rho}{\pd r}},
\end{equation}
or
\begin{equation}
\kappa
\sim
\frac{B}{C} \sqrt{-\frac{\pd P}{\pd r}\frac{\pd \rho}{\pd r}} \frac{r}{ \rho c_\mathrm{s}}.
\end{equation}
To obtain a form similar to Equation~(\ref{eq:vbal}), we
 note that
$\pd \rho/\pd r$ is directly related to the density contrast $\delta \rho$
between the plumes and the background flow in the framework of a
mixing-length approach (if we discount compressibility effects
as in \citealt{duffell_16} and \citealt{paxton_17}).
Using $\kappa=\delta v^2/c_\mathrm{s}^2$ and discarding non-dimensional
coefficients of order unity, we thus obtain:
\begin{equation}
\delta v^2
\sim
\frac{r c_\mathrm{s}}{\rho}
\sqrt{-\frac{\pd P}{\pd r}\frac{\pd \rho}{\pd r}}
\sim
\frac{r c_\mathrm{s}}{\rho }
\sqrt{ \frac{\rho g_\mathrm{eff}\, \delta \rho}{\Lambda}}
\sim
\frac{r c_\mathrm{s}}{\rho}
\sqrt{\frac{\rho g_\mathrm{eff} c_\mathrm{s}\,\delta \rho}{r\, \delta v}},
\end{equation}
or
\begin{equation}
\delta v^{5/2}
\sim
c_\mathrm{s}^{3/2}
\sqrt{ g_\mathrm{eff} r \frac{\delta \rho}{\rho}}.
\end{equation}
The resulting plume velocity $\delta v \sim \delta
v_\mathrm{bal}^{2/5} c_\mathrm{s}^{3/5}$ is somewhere in between the
equilibrium velocity $\delta v_\mathrm{bal}$ from
Equation~(\ref{eq:vbal}) and the sound speed. Considering that the
plume velocities are typically of a similar order of magnitude of
$c_\mathrm{s}$ anyway, this may not have a major effect. It is
probably still advisable to modify the turbulent damping term
in Equation~(\ref{eq:vconv}) to better capture the interplay between
buoyancy and drag forces. 
This could easily be achieved by  modifying
the  source and sink terms to reflect the work expended
against the drag force.
The sink term then needs to be proportional to
$\delta v^3/\lambda$; note that  such a cubic
dissipation term is also
well established in 1D turbulence models for subsonic stellar
convection \citep{kuhfuss_86,wuchterl_98}.
The source term also needs to be modified:
While the term $B \omega_\mathrm{RT}\kappa$ given in
\citet{duffell_16} correctly reproduces the exponential
growth during the initial phase of the
Rayleigh-Taylor instability, the instability enters
a different regime once elongated plumes form.
The growth rate of the kinetic energy is then
given by the product of the velocity perturbations
$\delta v$ and the force felt by plumes with density contrast
$\delta \rho$,\footnote{
Note that we again use $\delta \rho
=\frac{\pd \rho}{\pd r} \Lambda$ here, i.e.\ we do still
retain the incomressible approximation and assume
that the mixing length is still given by
Equation~(\ref{eq:lambda}). These assumptions will
be revisited below.}
\begin{equation}
   \left( \frac{\pd \rho \kappa c_s^2}
    {\pd t}\right)_\mathrm{source}=
\delta v\, g_\mathrm{eff} \frac{\delta \rho}{\rho}
=
\sqrt{\kappa} c_\mathrm{s}
\Lambda \omega_\mathrm{RT}^2
=
\kappa r c_\mathrm{s}
\omega_\mathrm{RT}^2.
\end{equation}
Since the quantity
$\rho \kappa$ rather than the turbulent kinetic
energy density is evolved in the
model of \citet{duffell_16}, the
modified evolution equation become
\begin{eqnarray}
\label{eq:vconv_new}
& &
\frac{\pd \rho \kappa}{\pd t}
+
\frac{1}{r^2}\frac{\pd}{\pd r}
\left[
r^2(\rho \kappa v_r-C \kappa c_\mathrm{s} r \frac{\pd \rho \kappa}{\pd r})
\right ]
=
\\
& &
\quad
(A+B \kappa) 
\frac{r}{\rho c_\mathrm{s}}
\times
\max (0,-\frac{\pd P}{\pd r}\frac{\pd \rho}{\pd r})
-D \kappa^{3/2} \rho c_\mathrm{s} \lambda^{-1},
\nonumber
\end{eqnarray}
where an appropriate approximation for the effective volume-to-surface
ratio $\lambda$ of the plumes is needed. This would give the desired
balance condition provided that local gradients can indeed be used to
estimate the density contrast.

We next examine whether local gradients can be used to obtain the
density contrast of the Rayleigh-Taylor plumes and the turbulent
transport of different species in the spirit of a mixing-length 
approach. To this end, we first compare (Figure~\ref{fig:dflukt}) the RMS fluctuations of the
density from our 3D model to the assumption that
the density contrast $\delta  \rho$ merely depends on the mixing
length and the density gradient (cp.\ Equation~(13) in \citealt{duffell_16}),
\begin{equation}
\label{eq:drho_ic}
\frac{\delta \rho}{\rho}
=C \Lambda \frac{\pd \ln \rho}{\pd r},
\end{equation}
and also to the more usual estimate from stellar mixing-length theory that
takes compressibility effects into account,
\begin{equation}
\label{eq:drho_c}
\frac{\delta \rho}{\rho}
=C \Lambda \left(\frac{\pd \ln \rho}{\pd r}-\frac{1}{\Gamma}\frac{\pd \ln P}{\pd r}\right).
\end{equation}
Here, the mixing $\Lambda$ is computed according to Equation~(\ref{eq:lambda})
using the RMS velocity fluctuations from the 3D model.
It should be noted that we
show $\sqrt{\delta \rho/\rho}$ instead of the density
contrast $\delta \rho/\rho$, in Figure~\ref{fig:dflukt}, since
the former determines the equilibrium velocity of the plumes
and is therefore a more appropriate metric for evaluating
an effective 1D turbulence model.

Neither the compressible nor the incompressible local approximation
fares very well at predicting the density contrast. Even a
recalibration of the proportionality factor $C$ would do little to
remedy this.  The local estimate for the density contrast becomes
especially problematic at later times (middle and bottom panel of
Figure~\ref{fig:dflukt}) when the reverse shock from the C/He
interface has formed and propagates deeper into the O/Ne/Mg/C shell.
However, Equation~(\ref{eq:drho_c}) for the compressible case at least
correctly predicts the sign of the density contrast in most regions
and appears to be a sufficiently good approximation to obtain the
plume velocity within a factor of a few.

Finally, we compare the mixing-length estimate for the partial mass
flux of species $i$,
\begin{equation}
\label{eq:fmix}
\mathcal{F}_\mathrm{i}
=4 \pi r^2 \langle \rho\rangle \delta v \Lambda \frac{\pd X_i}{\pd r},
\end{equation}
to th the turbulent partial mass flux of species $i$ in
the 3D simulation,
\begin{equation}
\mathcal{F}_\mathrm{i}
=\oint r^2 \left[(\rho-\langle \rho\rangle) (X_i-\langle X_i)\rangle (v_r-\langle v_r\rangle)-\langle X_i\rangle \langle \rho \rangle \langle v_r \rangle \right] \, \ud \Omega,
\end{equation}
where $X_i$ denotes the mass fraction of species $i$ and angled brackets denote
spherical Favre averages.\footnote{In other words, volume-weighted spherical
averages are used for the density, and density-weighted averages are used for other
quantities. The comparison between the 3D results and the mixing-length estimates
is not changed substantially by using volume-weighted averages instead.}
Results for $^{16}\mathrm{O}$ and the most abundant representative
neutron-rich iron group nucleus, $^{60}\mathrm{Fe}$, are shown
in Figure~\ref{fig:xflux}.

Again, the agreement between the diffusive approximation and the
actual turbulent fluxes is not too convincing. For $^{16}\mathrm{O}$,
there is rough agreement within a factor of two at early times, but
this agreement subsequently deteriorates.  We even encounter
situations where the actual turbulent flux points in the \emph{same}
direction as the gradient of the mass fluxes, i.e., a mass flux, which
is anti-diffusive, for example in the region between enclosed masses
of $1.44 M_\odot$ and $1.45 M_\odot$ during the later phases. That
such non-diffusive mixing occurs in both the 3D model and the 2D model
was in fact already evident from Figure~\ref{fig:mix_mass}, which
showed a considerably higher mass fraction of iron group elements at
the base of the He envelope than deeper down within the O/Ne/Mg/C
shell.  The non-diffusive nature of the mixing is also not unexpected
from the morphology of the Rayleigh-Taylor instability in supernova
simulations; very often one finds strongly elongated plumes
originating from deep in the supernova that have traversed overlying
shells without substantial small-scale mixing so that the initial
layering is partially inverted rather than erased by diffusive mixing.

Taken together, our analysis of the turbulent fluctuations and fluxes
in the 3D model does not provide convincing, positive justification
for an effective 1D treatment of Rayleigh-Taylor mixing in
core-collapse supernovae. This does not imply, however, that such an
approach is invalid as an approximation.  The comparison with 3D models of
\citet{wongwathanarat_15} in the work of \citet{paxton_17}
demonstrated that their 1D turbulence model can at least passably
mimick the end result at shock breakout for some progenitors.  More
work is needed to ascertain whether it does so accidentally, or
whether there is a deeper reason that allows it to approximately
reproduce the gross features of mixing on a global scale even though
some of its basic equations may not be very good approximations
locally.

\section{Applications to Observed Double Neutron Star Systems}\label{sec:applications}
\begin{figure}
\includegraphics[angle=-90,width=\linewidth]{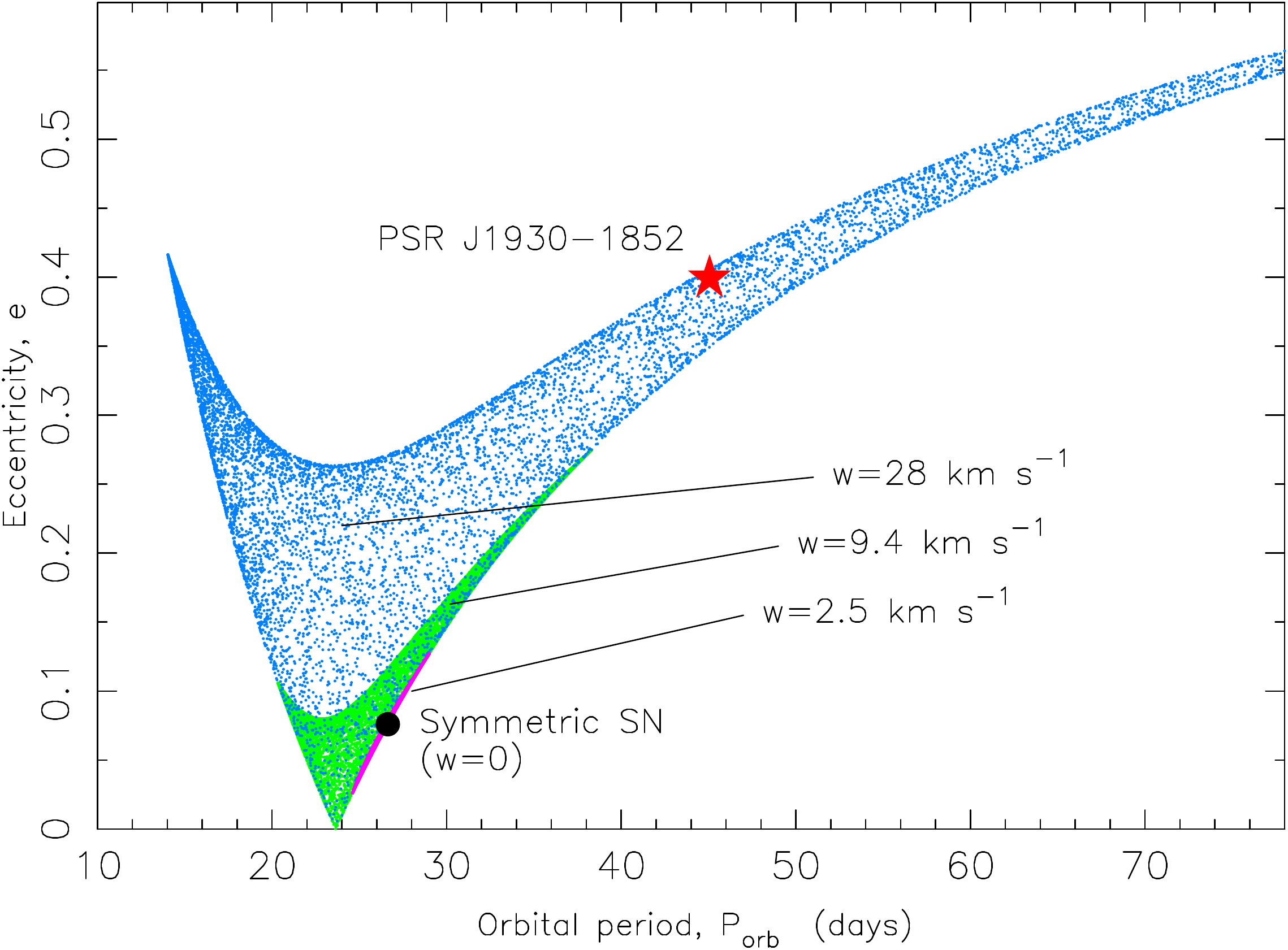}
\caption{Simulation of post-SN DNS systems in the orbital period--eccentricity plane, based on the properties of our ultra-stripped progenitor star and the estimated kick velocity magnitudes (colour coded: $w=2.5$, $9.4$ and $28\,{\rm km\,s}^{-1}$) obtained from our three different SN models: s2.8-D2-b, s2.8-3D and s2.8-2D-a, respectively.
The red star shows the properties of the widest known DNS system, PSR~J1930$-$1852 \citep{srm+15}.
\label{fig:post-SN-orbit}}
\end{figure}
Based on the detailed pre-SN evolution of the helium star--NS binary using a binary stellar evolution code \citep{tauris_15}, in this work extended up to collapse using
the \textsc{Kepler} code,
and on our models of the ensuing supernova explosion,
we can calculate the post-SN orbital properties of the resulting binary system. 

Originally, the binary system consisted of a $2.8\,M_{\odot}$ helium
star--NS binary with an orbital period of $20\,\mathrm{d}$.  At the
onset of oxygen burning, the helium star has become a detached star
with a total mass of $1.72\,M_{\odot}$ (a $1.47\,M_{\odot}$ metal core
and an envelope mass of $\mathord{\sim} 0.25\,M_{\odot}$, of which
$\mathord{\sim}0.22\,M_{\odot}$ is helium).  The orbital period at
this point was $19.4\,\mathrm{d}$.  It was assumed that the NS
companion had an initial mass of $1.35\,M_{\odot}$ and thus we also
take this value to be its final mass, since it only accreted
$3.2\times 10^{-4}\,M_{\odot}$ during the accretion phase
\citep{tauris_15}.

In this work (see Section~\ref{subsec:progenitor}), we modelled the
final stages of nuclear burning using the \textsc{Kepler} code and
found a strong off-centre silicon deflagration flash which ejected all
but $0.02\,M_{\odot}$ of the helium envelope, about $78\,\mathrm{d}$
prior to core collapse.  At the time of the SN explosion, the total
mass of the ultra-stripped star is $1.48\,M_{\odot}$.  Depending on
the details of the envelope ejection from the silicon flash, we find
that the pre-SN orbital period is between $22.8$ and
$23.1\,\mathrm{d}$.  In Section~\ref{subsec:kick-spin}, we presented
the resulting NS kick velocity from our modelling, ranging between
$2.5$ to $28\,{\rm km\,s}^{-1}$, from which we can calculate the
post-SN orbital properties of the resulting DNS system.  The
gravitational mass of the newly formed NS is $1.28\,M_{\odot}$, as
mentioned earlier.

In Figure~\ref{fig:post-SN-orbit}, we show the distribution of
possible DNS systems following the ultra-stripped SN explosion of our
star. Similarly to the method outlined in, e.g., \citet{ft14}, we
simulate $10,000$ SN explosions using Monte Carlo techniques to obtain
a random kick direction in each event, assuming an isotropic
distribution and a circular pre-SN orbit.  The resulting distributions
of DNS orbits are seen to cover a fairly large area in the orbital
period--eccentricity plane, depending on the kick magnitude ($w=2.5$,
$9.4$ or $28\,{\rm km\,s}^{-1}$).  We note that, by coincidence, we
are able to reproduce the properties of the wide-orbit DNS system
PSR~J1930$-$1852 \citep{srm+15}.  This demonstrates that
ultra-stripped SN can indeed be responsible for even the widest orbit
DNS system known.  For a full range of solutions to PSR~J1930$-$1852,
see the appendix of \citet{tauris_17}.

\section{Conclusions}
\label{sec:conclusions}
In this work, we performed the first detailed modelling of the SN
explosion of an ultra-stripped star produced from a binary stellar
evolution code. This means that different from previous work
\citep{suwa_15}, we actually calculate the mass loss and orbital
evolution of the supernova progenitor model up to collapse.  Our
initial model is a $2.8\,M_{\odot}$ helium star--NS binary with an
orbital period of 20~d. At the onset of core collapse, the donor star
has been reduced to an ultra-stripped star with a total mass of
$\mathord{\sim}1.48\,M_{\odot}$ as a result of Case~BC RLO followed by
a silicon deflagration flash after detachment.  Hence, in this model
the helium envelope is removed in a three-step process: first by the
stellar wind mass loss prior to Case~BC RLO, then as a result of mass
transfer to the NS companion, and finally, after detachment, via a
silicon deflagration flash. The result is an ultra-stripped SN with a
tiny envelope containing only $0.02\,M_{\odot}$ of
helium. Interestingly, the almost complete expulsion of the envelope
by the silicon flash occurs less than $100 \, \mathrm{d}$ before
explosion, which implies that the ensuing supernova may exhibit
interaction with circumstellar material (CSM).

We performed one 3D simulation and two 2D simulations of the SN
explosion using the neutrino hydrodynamics code
\textsc{CoCoNuT-FMT}. The outcome of these simulations is broadly in
agreement with previous work of \citet{suwa_15} who investigated the
explosion of bare C/O cores in 2D.  We find rapid shock revival and
small explosion energies of order $\mathord{\sim}10^{50} \,
\mathrm{erg}$.  The explosion dynamics is reminiscent of explosions of
low-mass iron core progenitors in the single star channel
\citep{mueller_12b,melson_15a,mueller_16b,radice_17}.  It is also
similar to ECSNe, though less extreme in the sense that shock
expansion is sufficiently slow to allow the development of the
low-mode asymmetries that are absent in ECSN models
\citep{wanajo_11,gessner_18}. Our results are also compatible with a
modest boost to the explosion energy due 3D turbulence as found by
\citet{melson_15a} for a low-mass iron core model.

By the end of the simulations, accretion onto the PNS has already
stopped, and the neutrino-driven wind emerges, so that an we can put
an upper limit of $1.28 \, M_\odot$ on the gravitational NS mass for
the current progenitor.  We see evidence for small NS kick velocities
between $2.5$ and $28\,{\rm km\,s}^{-1}$ for the ultra-stripped SN, in
agreement with \citet{suwa_15}. Our results confirm current ideas that
ultra-stripped core-collapse SNe of small iron cores lead to small
kicks \citep{tauris_15,tauris_17}. We do not find any evidence for a
spin-kick alignment in such SNe, as opposed to some empirical evidence
presented for isolated radio pulsars \citep[][and references
  therein]{noutsos_13}.

Based on the calculated pre-SN orbital properties and the donor star
mass at the onset of core collapse, the ejecta mass, the kick velocity
and the final gravitational mass of the resulting NS, we simulated the
post-SN binary systems using Monte Carlo techniques. We are able to
reproduce the properties of the widest orbit DNS system known
\citep[PSR~J1930$-$1852,][]{srm+15} which has an orbital period of
45~d. This suggests that ultra-stripped SNe are not only relevant for
tight-obit DNS systems but even apply for such wide-orbit systems. The
total mass of our DNS system is $2.63\,M_{\odot}$ compared to
$2.59\,M_{\odot}$ measured for PSR~J1930$-$1852 \citep{srm+15}.

As a preliminary step towards the calculation of observable
signatures, we followed the evolution of the 3D SN model and one 2D
model up to shock breakout.  Mixing driven by the Rayleigh-Taylor
instability develops at the base of the O/Ne shell and at the base of
the He envelope.  Isolated dense plumes of iron group material make it
roughly half way through the He shell.  The result is a considerable
presence of iron group elements (mass fraction $\mathord{\sim}0.1$)
throughout the outer layers of the ejecta.

As it has been recently proposed that Rayleigh-Taylor mixing in
supernova envelopes is amenable to a simple mixing-length treatment
(MLT) \citep{duffell_16,paxton_17}, we quantitatively investigated
turbulent velocity fluctuations and fluxes in our 3D model to check
the validity of such an MLT approach. Our simulation suggests that the
plume velocities are well described by balance between buoyancy and
drag forces, which could be captured in a modified 1D turbulence
model.  We found, however, that the turbulent fluxes cannot be well
approximated by diffusive fluxes.  Further work is necessary to
determine to what extent an effective 1D treatment of the
Rayleigh-Taylor instability is possible and can at least furnish a
rough global approximation for the mixing in supernova envelopes.

The investigation presented here is the first attempt to model the
evolution leading to the second SN in forming a DNS system by
combining detailed binary stellar evolution and state-of-the art
multi-dimensional SN modelling. In the future, we plan to investigate
such models for ultra-stripped SNe in tighter systems leading to
post-SN DNS systems that will merge within a Hubble time, thus leading
to systems similar to those detected by LIGO-Virgo. Furthermore, we
will attempt to explode ultra-stripped stars with more massive iron
cores to test the hypothesis of a correlation between NS mass and kick
velocity \citep{tauris_17}, and to determine whether one can obtain
  substantially more energetic explosion with higher nickel mass that
  are more similar to observationally inferred values
  \citep{drout_13}.  Follow-up work is also needed on the observable
signatures of ultra-stripped supernovae. It will be necessary to put
radiative transfer calculations of such events \citep{moriya_17} on a
more solid basis by incorporating the results of 3D simulations of
mixing instabilities as presented in this work.  Moreover, the
intriguing possibility of CSM interaction for explosions of
ultra-stripped progenitors needs to be investigated further.

\section*{Acknowledgements}
This work was supported by the Australian Research Council through an
ARC Future Fellowships FT160100035 (BM) and Future Fellowship
FT120100363 (AH) and by STFC grant ST/P000312/1 (SAS, BM).  This
material is based upon work supported by the National Science
Foundation under Grant No. PHY-1430152 (JINA Center for the Evolution
of the Elements).  This research was undertaken with the assistance of
resources from the National Computational Infrastructure (NCI), which
is supported by the Australian Government and was supported by
resources provided by the Pawsey Supercomputing Centre with funding
from the Australian Government and the Government of Western
Australia.  This work used the DiRAC Data Centric system at Durham
University, operated by the Institute for Computational Cosmology on
behalf of the STFC DiRAC HPC Facility (\url{www.dirac.ac.uk}); this
equipment was funded by a BIS National E-infrastructure capital grant
ST/K00042X/1, STFC capital grant ST/K00087X/1, DiRAC Operations grant
ST/K003267/1 and Durham University. DiRAC is part of the UK National
E-Infrastructure.

\bibliography{paper}

\label{lastpage}

\end{document}